\documentclass[aps,prl,reprint,superscriptaddress]{revtex4-2}
\usepackage{amsfonts}
\usepackage{amsmath}
\usepackage{amssymb}
\usepackage{graphicx}
\usepackage{epstopdf}
\usepackage{color}
\usepackage{bbold}
\usepackage[colorlinks, linkcolor=blue, citecolor=blue, urlcolor=blue]{hyperref}

\renewcommand{\vec}[1]{\boldsymbol{#1}}

\begin{document}
\title{A Theory of Anisotropic Magnetoresistance in Altermagnets and Its Applications}
%\title{Electric readout of the Néel vector in an altermagnet}

\author{Xian-Peng Zhang}

\affiliation{Centre for Quantum Physics, Key Laboratory of Advanced Optoelectronic Quantum Architecture and Measurement (MOE), School of Physics, Beijing Institute of Technology, Beijing, 100081, China}
\affiliation{Department of Physics, Hong Kong University of Science and Technology, Clear Water Bay, Hong Kong, China}
\affiliation{International Center for Quantum Materials, Beijing Institute of Technology, Zhuhai, 519000, China}

\author{Run-Wu Zhang}

\affiliation{Centre for Quantum Physics, Key Laboratory of Advanced Optoelectronic Quantum Architecture and Measurement (MOE), School of Physics, Beijing Institute of Technology, Beijing, 100081, China}

\author{Xiaolong Fan}
\affiliation{Key Laboratory of Magnetism and Magnetic Materials (MOE), School of Physics Science and Technology,
Lanzhou University, Lanzhou 730000, China}

\author{Wanxiang Feng}
\email{wxfeng@bit.edu.cn}
\affiliation{Centre for Quantum Physics, Key Laboratory of Advanced Optoelectronic Quantum Architecture and Measurement (MOE), School of Physics, Beijing Institute of Technology, Beijing, 100081, China}
\affiliation{International Center for Quantum Materials, Beijing Institute of Technology, Zhuhai, 519000, China}

\author{Xiangrong Wang}
\email{phxwan@ust.hk}
\affiliation{Department of Physics, Hong Kong University of Science and Technology, Clear Water Bay, Hong Kong, China}

\author{Yugui Yao}
\email{ygyao@bit.edu.cn}
\affiliation{Centre for Quantum Physics, Key Laboratory of Advanced Optoelectronic Quantum Architecture and Measurement (MOE), School of Physics, Beijing Institute of Technology, Beijing, 100081, China}

\affiliation{International Center for Quantum Materials, Beijing Institute of Technology, Zhuhai, 519000, China}

\begin{abstract}
Altermagnets, a newly discovered class of magnets, integrate the advantages of both ferromagnets and antiferromagnets, such as enabling anomalous transport without stray fields and supporting ultrafast spin dynamics, offering exciting opportunities for spintronics. A key challenge in altermagnetic spintronics is the efficient reading and writing of information by switching the Néel vector orientations to represent binary “0” and “1”.  Here, we develop a microscopic theory of the magnetoresistance effect in altermagnets and propose that magnetoresistance anisotropy can serve as an effective mechanism for the electrical readout of the Néel vector. Our theory describes a two-step charge-spin-charge conversion process governed by the interplay between spin splitting and spin Hall effects: a longitudinal electric field induces transverse drift spin currents, which induce significant spin accumulation at the boundaries, generating a diffusive spin current that is converted back into a longitudinal charge current. By switching the Néel vector, a substantial change in magnetoresistance, akin to giant magnetoresistance in ferromagnets, is realized, enabling an electrically readable altermagnetic memory. Our microscopic theory provides deeper insights into the fundamental physics of the magnetoresistance effect in altermagnets and offers valuable guidance for designing next-generation ultradense and ultrafast spintronic devices based on altermagnetism.
\end{abstract}

\maketitle

\textit{Introduction --} Altermagnets (AMs) have recently been identified as a third class of magnetic phases, distinct from conventional ferromagnets and antiferromagnets~\cite{libor2022beyond,libor2024emerging,bai2024altermagnetism}. Like antiferromagnets, AMs exhibit compensating antiparallel magnetic order in real space, resulting in vanishing macroscopic net magnetization. However, unlike conventional antiferromagnets, where different magnetic sublattices are related by time-reversal symmetry combined with spatial inversion or translation, AMs are instead characterized by a symmetry operation that couples time reversal with crystal rotation. This unique symmetry breaks global Kramers spin degeneracy, leading to alternating spin-split electronic bands along specific paths in reciprocal space~\cite{Fedchenko2024,Reimers2024,Krempasky2024}. The resulting spin order parameter in AMs gives rise to various exotic transport phenomena, including the nonrelativistic spin-polarized currents~\cite{gonzalez2021efficient,bai2022observation},  anomalous Hall effect~\cite{libor2020AHE,feng2022anomalous}, and magneto-optic effects~\cite{zhou2021crystal,Hariki2024,Gray2024}, all previously considered hallmarks of FMs. With their zero stray field and terahertz spin dynamics inherited from antiferromagnets, AMs hold great promise for next-generation spintronic applications, such as ultradense and ultrafast nonvolatile magnetic random-access memory.

Due to the absence of net magnetization, reading and writing information in altermagnetic memory—where binary “0” and “1” are represented by opposite Néel vector orientations—remains a significant challenge. One potential readout mechanism is the anomalous Hall effect, as the Hall resistivity reverses sign when the Néel vector in AMs flips~\cite{libor2020AHE,feng2022anomalous,han2024electrical}. However, being a transverse transport phenomenon, the anomalous Hall effect requires at least four electrical contacts for measurement, complicating device fabrication. A more straightforward alternative is to utilize the longitudinal magnetoresistance (MR) effect, in which the charge current flows parallel to the applied electric field. Importantly, MR-based technologies are highly compatible with existing CMOS processes and are already widely used in commercial magnetic random-access memory. Drawing from established models in ferromagnetic systems, theoretical frameworks for giant and tunneling MR effects have been proposed for AMs~\cite{libor2022giant,Shao2021}. Experimentally, various MR effects have recently been reported in altermagnetic materials such as RuO$_2$~\cite{Chen2024}, MnTe~\cite{Betancourt2024}, CrSb~\cite{Peng2024}, and MnF$_2$~\cite{Liu2024}. Nevertheless, the microscopic origins of the MR effect in AMs—and its anisotropy with respect to the Néel vector orientation, especially from the standpoint of spin dynamics—remain largely unexplored and are not yet fully understood. 

\begin{figure}[t!]
\begin{center}
\includegraphics[width=0.48\textwidth]{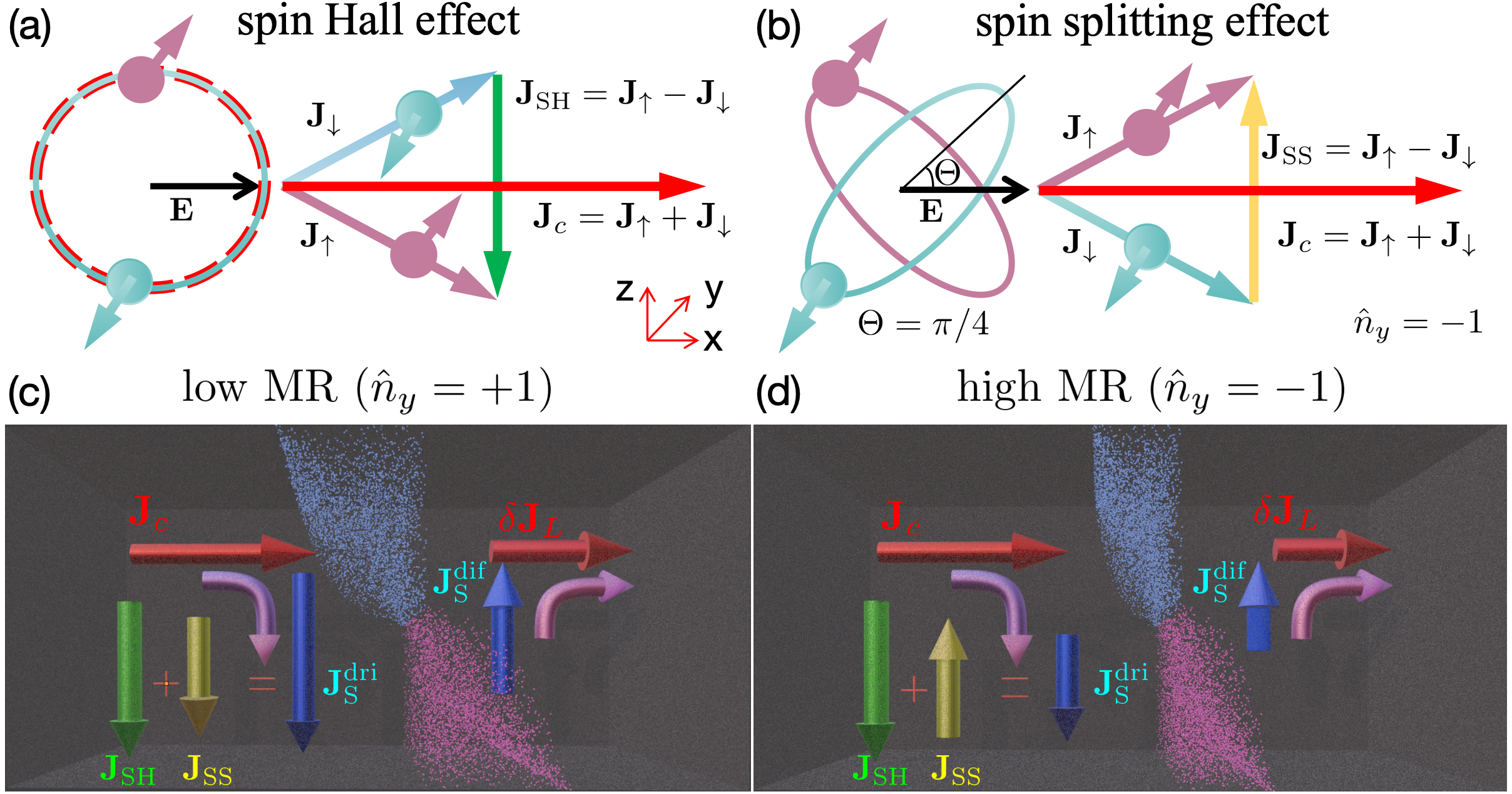} 
\end{center}
\caption{(a,b) Schematics of transverse spin currents arising from the SHE and SSE, which originate from isotropic and anisotropic spin-splitting bands, respectively. $\vec{E}$ denotes the electric field applied along the $x$-axis; $\vec{J}_c$ is the longitudinal charge current, decomposed into spin-up ($\vec{J}_\uparrow$) and spin-down ($\vec{J}_\downarrow$) components. $\vec{J}_{\text{SH}}$ and $\vec{J}_{\text{SS}}$ are the transverse spin currents along the $z$-axis, generated by the SHE and SSE, respectively. The spin polarization direction of the SSE is pinned to the Néel vector, and is   schematically set to be antiparallel to that of the SHE,  i.e., along the $y$-axis with $n_y =\sin\alpha =-1$.	(c,d) The interplay between SHE and SSE leads to low and high MR states when the Néel vector is parallel ($n_y =\sin\alpha=+1$) or antiparallel ($n_y =\sin\alpha= -1$) to the spin polarization of the SHE-induced spin current, respectively. The MR effect originates from a two-step charge-spin-charge conversion process. In the first step, longitudinal charge currents ($\vec{J}_c$) are converted into transverse drift spin currents ($\vec{J}^{\text{dri}}_{\text{S}}$) via the combined action of SHE and SSE, generating  significant spin accumulation $\mu^{y}_{s}(z)$ along the $z$-direction. In the second step, the diffusive spin currents ($\vec{J}^{\text{dif}}_{\text{S}}$) reflected from the sample boundaries are converted back into charge currents ($\delta\vec{J}_{L}$) through the respective inverse processes. Parallel (antiparallel) alignment yields a larger (smaller) drift spin current, $\vec{J}^{\text{dri}}_{\text{S}} = \vec{J}_{\text{SH}} + \vec{J}_{\text{SS}}$ ($\vec{J}_{\text{SH}} - \vec{J}_{\text{SS}}$), resulting in a larger (smaller) reflected spin current that is converted back into charge current with higher (lower) efficiency, $\theta_{\text{SH}} + \theta_{\text{SS}}$ ($\theta_{\text{SH}} - \theta_{\text{SS}}$), thereby leading to low (high) MR.}
\label{STORY}
\end{figure}

In this Letter, we present a comprehensive theory that elucidates the physical origins of the MR effect and its anisotropy in AMs. Our approach focuses on microscopic spin dynamics, including spin flip and relaxation, extending beyond traditional ferromagnetic models.  In the non-relativistic limit, a drift spin current with polarization aligned to the Néel vector is induced perpendicular to the applied electric field, known as the spin splitting effect (SSE) [Fig.~\ref{STORY}(b)]. This spin current, arising from alternating spin-split bands, exhibits strong anisotropy with respect to the field direction. When spin-orbit coupling is introduced, an additional isotropic drift spin current emerges, also perpendicular to the applied field, corresponding to the spin Hall effect (SHE) [Fig.~\ref{STORY}(a)]. In real altermagnetic materials, both spin currents coexist, leading to spin accumulation at the sample boundaries, which, in diffusive regime, generates a diffusive spin current opposite to drift one and subsequently converts into a longitudinal charge current via inverse spin-splitting and inverse spin-Hall effects. We propose that this two-step charge-spin-charge conversion process constitutes the microscopic mechanism of the MR effect in AMs. Moreover, switching the Néel vector orientation significantly alters the MR, with distinct low- and high-resistance states, which are crucial for realizing electrically readable altermagnetic memory devices [Figs.~\ref{STORY}(c) and~\ref{STORY}(d)].  The MR effect and its anisotropy proposed in this work require only a single altermagnetic material, eliminating the need for multilayer structures, such as those combining with nonmagnetic metals or insulators, used in giant and tunneling MR effects. This significantly reduces the challenges associated with experimental thin-film fabrication, particularly the stringent requirement for ultra-clean interfaces. As such, our work provides valuable guidance for designing next-generation, high-performance spintronic devices based on altermagnetism.

\textit{Hybrid system's Hamiltonian --} We start by considering the Hamiltonian of a hybrid system that captures the interaction between free electrons and local spin moments: $H=H_{e}+H_{m}+V_{em}$. The first term, $H_{e}$, describes the $i$-th free electron moving in the $xy$-plane (with finite size along the $z$-axis), and is given by $H_{e}=p^2_i/2m+\lambda[\sin2\Theta  p_i^xp_i^y+\cos2\Theta(p^x_ip^x_i-p^y_ip^y_i)/2]\sigma^{\vec{n}}_i-(l^2_{\text{so}}/\hbar)\vec{\sigma}_i \cdot [\vec{p}_i \times \vec{\nabla} V_{\text{so}}(\vec{r}_i)]$, where $\vec{r}_i$ and $\vec{p}_i$ denote the position and momentum, respectively, and $\sigma^{\vec{n}}_i$ is the spin polarization's direction, defined as the dot product between the Pauli matrix vector $\vec{\sigma}_i=[\sigma_i^x,\sigma_i^y,\sigma_i^z]$  and the Néel vector $\vec{n}$, which can be tuned by external magnetic fields~\cite{Betancourt2023,Reichlova2024,feng2022anomalous} or electric fields~\cite{wadley2016electrical,Godinho2018,han2024electrical}. Here, $\lambda$ characterizes the spin splitting strength in altermagnets, and $\Theta$ denotes the orientation of spin-split anisotropy with $x$ axis as indicated by Fig. \ref{STORY}(b)~\cite{sun2023dndreev,beenakker2024phase}. $V_{\text{so}}(\vec{r}_i)$ represents the spin-orbit potential, and the constant $l_{\text{so}} = \hbar / (2mc)$ (where $m$ is the electron mass and $c$ is the speed of light in vacuum)~\cite{winkler2003spin,messiah2014quantum}. The second term, $H_{m}$, describes the exchange interactions between local spin moments and is expressed as $H_{m}=- \frac{1}{2}J_{\imath n,\jmath m}\mathbf{S}_{\imath n}\cdot \mathbf{S}_{\jmath m}$, where $\mathbf{S}_{\imath n}=[\text{S}^x_{\imath n},\text{S}^y_{\imath n},\text{S}^z_{\imath n}]$ is the spin-S operator at position $\vec{R}_{\imath n}$ (with $\imath = a$ or $b$ indicating the sublattice in unit cell $n$), and the negative exchange energy $J_{\imath n,\jmath m} < 0$ enforces an antiparallel spin configuration. All spins within the same sublattice are assumed to be identical, as reflected by the spin expectation value $\langle \text{S}_{\imath}^{\nu} \rangle = \langle \text{S}_{\imath n}^{\nu} \rangle$ and the spin-spin correlation function $\langle \text{S}_{\imath}^{\nu} \text{S}_{\imath}^{\nu} \rangle = \langle \text{S}_{\imath n}^{\nu} \text{S}_{\imath n}^{\nu} \rangle$ for $\nu = \{x, y, z\}$. The last term, $V_{em}$, represents the scattering potential experienced by electrons due to local spin moments, and is given by~\cite{zhang2019theory,zhang2023extrinsic,zhang2022microscopic}
\begin{equation} \label{spin-exchange}
	V_{em} = -\mathcal{J} \vec{\sigma}_i \cdot \mathbf{S}_{\imath n} \delta(\vec{r}_i - \vec{R}_{\imath n}),
\end{equation}
where $\mathcal{J}$ is the spin-exchange coupling (SEC) constant between free electrons and local spin moments, assumed to be identical for both sublattices. In our calculations, the SEC is effectively parameterized as $\mathcal{J}n_s$, where $n_s$ denotes the density of local moments per sublattice. The value of $\mathcal{J}n_s$ is estimated to be on the order of meV~\footnote{The SEC for Co adatoms on the Cu(100) surface has been theoretically predicted to reach a substantial ferromagnetic interaction of approximately $\mathcal{J}n_s \simeq 350$ meV~\cite{wahl2007exchange}. Although experimental data for SEC in the altermagnets CrSb and RuO$_2$ are currently unavailable, we consider $\mathcal{J}n_s \sim$ meV to be a physically reasonable estimate.}.

\textit{Anisotropic spin diffusion equation --} The key role of the SEC, Eq.~\eqref{spin-exchange}, in the MR effect lies in its induction of anisotropic spin relaxation for free electrons. Based on the standard theory of open quantum systems~\cite{zhang2024microscopic,breuer2002theory}, the longitudinal and transverse spin relaxation times are given, respectively, by
\begin{align} \label{LongiSRT}
	\frac{1}{\tau_{\Vert}} &= \frac{1}{\tau_0}
	+\frac{\pi}{\hbar} n_s\nu_F\mathcal{J}^2 \beta\epsilon_L^{\imath}
	n_{B}(\epsilon^{\imath}_L)\left[\text{S}(\text{S}+1)-\langle \text{S}_{\imath}^{\Vert}\text{S}_{\imath}^{\Vert}\rangle - \langle \text{S}_{\imath}^{\parallel}\rangle\right],
\end{align}
\begin{align}  \label{TransSRT}
	\frac{1}{\tau_{\perp}} &= \frac{1}{2\tau_0} + \frac{1}{2\tau_{\Vert}}
	+ \frac{\pi}{\hbar} n_s \nu_F \mathcal{J}^2 \langle \text{S}_{\imath}^{\Vert} \text{S}_{\imath}^{\Vert} \rangle.
\end{align}
Here, the first term, $\tau_0^{-1}$, denotes the spin relaxation rate unrelated to local moments. It originates from the spin-orbit potential $V_{\text{so}}$ and is assumed to be isotropic and independent of temperature ($T$).  The second term in Eq.~\eqref{LongiSRT} corresponds to the spin-flip rate arising from magnon emission and absorption~\cite{zhang2024microscopic}, where $\nu_F$ is the density of states per spin at the Fermi level, $\mathrm{S}_{\imath}^{\Vert}$ denotes the spin component parallel to the Néel field direction, and $n_B(\epsilon^{\imath}_L) = 1/(e^{\beta\epsilon^{\imath}_L} - 1)$ is the Bose-Einstein distribution function with inverse temperature $\beta = 1/(k_BT)$.  The effective Larmor frequency is written as $\epsilon^{\imath}_{L} = -\sum_{\jmath n} J_{\imath n,\jmath m} \langle \text{S}_{\jmath n}^{\Vert} \rangle$. The third term in Eq.~\eqref{TransSRT} accounts for the spin dephasing rate arising from scattering processes, during which the electron’s spin accumulates a precession phase~\cite{zhang2019theory,zhang2024microscopic}. The anisotropic spin relaxation time induced by the SEC depends strongly on $T$ through $\langle \text{S}^{\Vert}_{\imath} \rangle$ and $\langle \text{S}^{\parallel}_{\imath} \text{S}^{\parallel}_{\imath} \rangle$, and is consequently responsible for the $T$-dependent behavior of MR effect.

Besides the SEC, the SSE and SHE also play crucial roles in altermagnetic MR through a two-step charge-spin-charge process. In the first step, longitudinal charge currents are converted into transverse drift spin currents [indicated by the curved arrows on the left sides of Figs.~\ref{STORY}(c) and~\ref{STORY}(d)]. The SSE generates a spin current whose polarization is aligned with the Néel vector [Fig.~\ref{STORY}(b)], i.e., $\vec{J}^{\nu}_{\mathrm{SS}} = \theta_{\text{SS}} n_{\nu} \hat{y} \times \sigma_{\text{D}} \vec{E}$~\cite{gonzalez2021efficient,liao2024separation,feng2024incommensurate,bai2022observation}, in contrast to the SHE-induced spin current, which is polarized perpendicular to the electron trajectory [Fig.~\ref{STORY}(a)], i.e., $\vec{J}^{\nu}_{\mathrm{SH}} = \theta_{\text{SH}} \hat{\nu} \times \sigma_{\text{D}} \vec{E}$~\cite{chen2013theory,maekawa2017spin,sinova2015spin}, for $\nu = \{x, y, z\}$.  Here, $\vec{E}$ denotes the electric field applied along the $x$-axis, $\sigma_{\text{D}}$ is the Drude conductivity, and $\theta_{\text{SS}}$ and $\theta_{\text{SH}}$ represent the charge-spin conversion ratio for the SSE and SHE, that rely on $\lambda$ and $V_{\text{so}}$, respectively [see Eqs.~\eqref{svfvf} and \eqref{svfv} in the End Matter].  The resulting drift spin current along the $z$-axis leads to significant spin accumulation near the sample boundaries, giving rise to a diffusive spin current described by $-\frac{\sigma_{\text{D}} }{2e} \partial_z \mu_{s}^{\nu}(z)$. In the steady state, the drift spin currents reaching the top ($z = d_N$) and bottom ($z = 0$) surfaces are fully canceled by the diffusive spin current, yielding the boundary condition:
\begin{align} \label{fvmdkmkd}
	-\frac{\sigma_{\text{D}} }{2e}\left. \partial _{z}\mu
	_{s}^{\nu }\right\vert _{z=0,d_{N}}-n_{\nu }\theta _{\mathrm{SS}}\sigma _{\text{D}}E-\delta _{\nu  y}\theta _{\mathrm{SH}}\sigma _{\text{D}}E=0,
\end{align}
where $d_N$ is the thickness of the altermagnetic thin film.

In the second step, the transverse diffusive spin current is converted back into a longitudinal charge current via the corresponding inverse effects [indicated by the curved arrows on the right sides of Figs.~\ref{STORY}(c) and~\ref{STORY}(d)]. The spin density $\mu_s^{\nu}(z)$ depends solely on the $z$-coordinate and satisfies the anisotropic spin diffusion equation~\cite{zhang2024microscopic}:
\begin{equation} \label{3.3}
	\partial_z^2 \mu_s^{\nu} = \ell_{\perp}^{-2} \delta_{\nu\kappa} \mu_s^{\kappa} + \left( \ell_{\parallel}^{-2} - \ell_{\perp}^{-2} \right) n_{\nu} n_{\kappa} \mu_s^{\kappa},
\end{equation}
where $\nu, \kappa = \{x, y, z\}$. The first and second terms describe the anisotropic spin relaxation of free electrons. The transverse and longitudinal spin diffusion lengths are given by $\ell_{\perp} = \sqrt{\mathcal{D} \tau_{\perp}}$ and $\ell_{\parallel} = \sqrt{\mathcal{D} \tau_{\parallel}}$, respectively, where $\mathcal{D}$ is the diffusion coefficient.

With the boundary condition [Eq.\eqref{fvmdkmkd}], the anisotropic spin diffusion equation [Eq.\eqref{3.3}] can be analytically solved (see detailed derivations in the Supplemental Material~\cite{SM}). The resulting expression for the longitudinal resistivity reads
\begin{align} \label{SL}
\Delta\rho _{\mathrm{L}}&\simeq \theta _{\mathrm{SH}%
}^{2}\Delta\rho _{1} \left( 1-n_{y}^{2}\right)-(\theta _{\mathrm{SH}}^{2}+\theta _{\mathrm{SS}}^{2})\Delta\rho _{0}\\
&-2\theta _{\mathrm{SH}
}^{}\theta _{\mathrm{SS}%
}^{}\Delta\rho _{0}n_y,\notag
\end{align}%
with
\begin{equation} \label{vvh0}
\frac{\Delta\rho _{0}}{\rho _{\mathrm{L0}}}=\frac{2\ell _{\Vert }}{d_{N}}\tanh \left( \frac{%
d_{N}}{2\ell _{\Vert }}\right),
\end{equation}%
\begin{equation} \label{vvh1}
\frac{\Delta\rho _{1}}{\rho _{\mathrm{L0}}}=\frac{2\ell _{\Vert  }}{d_{N}}\tanh \left( \frac{%
d_{N}}{2\ell _{\Vert  }}\right)- \frac{2\ell _{\perp  }}{d_{N}}\tanh \left( \frac{%
d_{N}}{2\ell _{\perp  }}\right),
\end{equation}%
where $\rho_{\mathrm{L0}} = 1 / \sigma_{\mathrm{D}}$ is the baseline longitudinal resistivity.  Equations~\eqref{SL}--\eqref{vvh1} represent the central results of this work. They reveal a novel anisotropy in the MR effect that depends on the angle between the Néel vector and $y$ axis ($n_y=\sin\alpha$) and provide microscopic insight into the MR mechanism via two distinct contributions: the magnon-induced spin-flip process [Eq.~\eqref{vvh0}] and the anisotropic spin relaxation of free electrons [Eq.~\eqref{vvh1}].

\textit{Anisotropic MR effect in altermagnets --} Next, we apply our MR theory to a realistic altermagnetic material, CrSb~\cite{Peng2024,zhou2025manipulation}, which features a high Néel temperature of $T_{\text{N}} \simeq 700$ K, making it a promising candidate for room-temperature applications. Although our analysis centers on CrSb, the theoretical framework developed here can be readily generalized to other altermagnets, such as RuO$_2$~\cite{feng2024incommensurate,bai2022observation} (see additional discussion in the End Matter).

\begin{figure}[t]
\begin{center}
\includegraphics[width=0.48\textwidth]{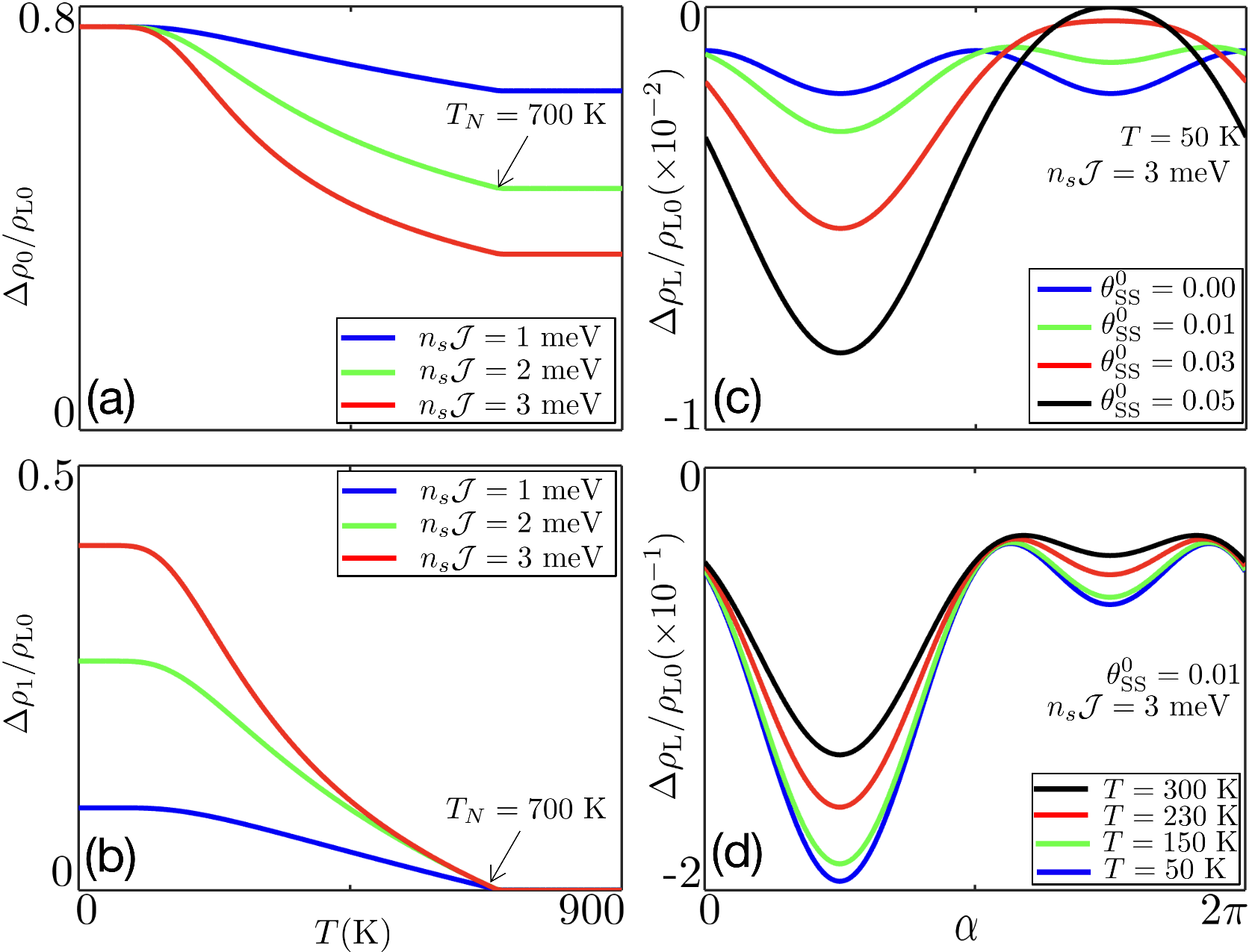} 
\end{center}
\caption{(a,b) The temperature ($T$) modulation of $\Delta\rho _{0}$ and $\Delta\rho _{1}$ for several values of $n_s\mathcal{J}$.  (c,d) Longitudinal resistivity as a function of the direction of the Néel vector ($n_y=\sin\alpha$) for different values of $\theta^0_{\text{SS}}$ and $T$, with $\Delta\rho_0/\Delta\rho_1\simeq 1.5$.  Other parameters: $\theta _{\mathrm{SH}}=0.05$, $\text{S}=3/2$, $\ell _{0}=3.0$ nm, and $d_{N}=5$ nm.}
\label{MMR}
\end{figure}

For simplicity, we first exclude the SSE by setting $\theta_{\mathrm{SS}} = 0$. In this case, the MR effect follows a cosine-square dependence,
$\theta_{\mathrm{SH}}^{2} \Delta\rho_1 (1 - n_y^2) = \theta_{\mathrm{SH}}^{2} \Delta\rho_1 \cos^2\alpha$, which exhibits symmetric behavior with respect to $+n_y$ and $-n_y$, and displays a $\pi$-periodic oscillation [Fig.~\ref{MMR}(c)]. The resulting $\pi$-periodic MR with respect to the Néel vector—rather than the net magnetization—resembles the anisotropic MR typically observed in antiferromagnets. This MR anisotropy originates from anisotropic spin relaxation, specifically the difference between $\ell_{\parallel}$ and $\ell_{\perp}$ [see Eq.\eqref{vvh1}]. To provide an intuitive explanation, we focus on the altermagnetic phase at sufficiently low temperature ($T \ll T_N$). When the spin polarization of free electrons is aligned with the Néel vector, spin relaxation is weak, such that $\tau^{-1}_{\parallel} \rightarrow \tau^{-1}_0$. As a result, more transverse spin current is reflected at the top and bottom boundaries and subsequently converted into a longitudinal charge current via the inverse SHE, leading to a low-resistance state.  In contrast, when the spin polarization is perpendicular to the Néel vector, the spin relaxation becomes significantly stronger, i.e., $\tau^{-1}_{\perp} = \tau^{-1}_0 + 2\Omega_0$, where $\Omega_0 = \frac{\pi}{\hbar} n_s \nu_F \mathcal{J}^2 \text{S}^2$. In this regime, the transverse spin current is more likely to decay during propagation and thus fails to be converted into a charge current via the inverse SHE, resulting in a high-resistance state. 

In the presence of the SSE, its interplay with the SHE significantly modifies the anisotropic MR effect. The influence of the SSE is primarily captured by the third term in Eq.~\eqref{SL}, which is linear in the Néel vector orientation ($n_y=\sin\alpha$). As a result, the MR effect transitions from the original $\pi$-periodicity (in $n_y^2$) to a $2\pi$-periodicity (in $n_y$) as $\theta^0_{\text{SS}}$ increases from 0 to 0.05 [Fig.~\ref{MMR}(c)]. Here, $\theta^0_{\text{SS}}$ denotes the SSE-induced charge-spin conversion ratio at zero temperature, and its temperature dependence is introduced through the expression $\theta_{\text{SS}} = \theta^0_{\text{SS}} (\langle \text{S}^{\Vert}_{b} \rangle - \langle \text{S}^{\Vert}_{a} \rangle)/(2\text{S})$. A quantitative analysis of this evolution requires accounting for the temperature dependence of both $\Delta\rho_0$ and $\Delta\rho_1$, which are plotted in Fig.~\ref{MMR}(a) and Fig.~\ref{MMR}(b), respectively. Notably, in the presence of SSE, the $2\pi$-periodic MR effect can be realized over a broad temperature range, as illustrated in Fig.~\ref{MMR}(d).

\textit{Electric readout of Néel vector --} The interplay between SHE and SSE is crucial for the electrical readout of altermagnetic memory devices. As revealed in Eq.~\eqref{SL}, an additional resistance contribution arising from the combined effect of SHE and SSE is linear in the Néel vector $n_y$, thereby enabling the distinction between opposite Néel vector orientations, corresponding to binary states “0” and “1”.  By considering the cases where the Néel vector is parallel ($n_y = +1$) or antiparallel ($n_y = -1$) to the polarization direction of the SHE-induced spin current [Figs.~\ref{STORY}(c) and~\ref{STORY}(d)], the charge current converted from the diffusive spin current via the inverse SHE and SSE is proportional to the spin accumulation at the edges, $\delta\vec{J}_{L} \propto (\theta_{\mathrm{SH}} + n_y \theta_{\mathrm{SS}})\left[ \mu^y_s(0) - \mu^y_s(d_N) \right]$, where $n_y = \pm1$ [see Eq. (S32) in SM~\cite{SM}]. In the parallel configuration ($n_y = +1$), the total drift spin current $\vec{J}_{\text{SH}} + \vec{J}_{\text{SS}}$ is enhanced, leading to a stronger diffusive spin current quantified by the edge spin accumulation $\mu^y_s(0) - \mu^y_s(d_N)$. This spin current is subsequently converted into a charge current with higher spin-to-charge conversion efficiency $\theta_{\mathrm{SH}} + \theta_{\mathrm{SS}}$, resulting in a low-resistance state.  Conversely, in the antiparallel configuration ($n_y = -1$), the drift spin current $\vec{J}_{\text{SH}} - \vec{J}_{\text{SS}}$ is reduced, resulting in a smaller diffusive spin current and a lower conversion efficiency $\theta_{\mathrm{SH}} - \theta_{\mathrm{SS}}$, leading to a high-resistance state. The resulting resistance difference between the two configurations is $4 \theta_{\mathrm{SH}} \theta_{\mathrm{SS}} \Delta\rho_0$, as indicated in Eq.~\eqref{SL}.

Next, we discuss in detail the change of MR with the switching of the Néel vector. As noted earlier, $\Delta\rho_0$ originates from magnon-induced spin-flip processes, as captured by the longitudinal relaxation rate $\tau^{-1}_{\Vert}$ [see Eqs.\eqref{LongiSRT} and\eqref{vvh0}], and its temperature dependence is shown in Fig.~\ref{MMR}(a) for different values of $n_s\mathcal{J}$. The isotropic spin diffusion length unrelated to local moments, $\ell_0 = \sqrt{\mathcal{D}\tau_0} = 3.0$ nm, is chosen to be comparable to the thickness of the CrSb film ($d_N = 5$ nm). As $\ell_{\Vert}/d_N$ increases, $\Delta\rho_0$ initially grows linearly as $\Delta\rho_0 \approx 2\ell_{\Vert}/d_N$, and gradually saturates towards unity, following the asymptotic relation $\Delta\rho_0 \approx 1 - d_N^2/(12\ell_{\Vert}^2)$. Notably, $\ell_{\Vert}$ evolves from $[\mathcal{D}/(\tau^{-1}_0 + \Omega_1)]^{1/2}$ in the high temperature nonmagnetic regime ($T > T_{\text{N}}$) to $\ell_0$ in the low-temperature altermagnetic regime ($T \ll T_{\text{N}}$), where
$\Omega_1 = \frac{2\pi}{3\hbar} n_s \nu_F \mathcal{J}^2 \text{S}(\text{S} + 1)$.  At temperatures above $T_{\text{N}}$, enhanced magnon emission and absorption processes (mediated by stronger $\mathcal{J}$) increase the spin-flip rate $\Omega_1 \propto \mathcal{J}^2$, thereby reducing the longitudinal spin diffusion length and suppressing $\Delta\rho_0$. In contrast, at low temperatures ($T \ll T_{\text{N}}$), the strong Néel field enforces antiparallel alignment of local spin moments, which suppresses magnon-related processes, causing $\Omega_1 \rightarrow 0$. As a result, $\Delta\rho_0$ becomes independent of $n_s\mathcal{J}$ and reaches a saturated value given by $\Delta\rho _{0} = \rho _{\text{L}0}\frac{2\ell _{0 }}{d_{N}}\tanh \left( \frac{d_{N}}{2\ell _{0 }}\right)$, which approaches $\rho_{\text{L}0}$ in the limit $\ell_0 / d_N \gg 1$.  Moreover, the charge-spin conversion ratio of the SSE can significantly exceed that of the SHE reported in conventional materials like Pt, i.e., $\theta_{\text{SS}} \gg \theta_{\text{SH}}$~\cite{gonzalez2021efficient}. Consequently, the MR in our system, which scales as $\propto \theta_{\text{SH}} \theta_{\text{SS}}$, can be substantially larger than conventional SHE-based MR, which scales as $\propto \theta_{\text{SH}}^2$.  This pronounced change in MR upon switching the Néel vector highlights the critical role of altermagnets in electrically readable memory applications, paving the way for next-generation spintronic devices~\cite{baltz2018antiferromagnetic}.

\textit{Conclusion --} We have developed a microscopic theory to elucidate the physical origins of MR in spin-orbit-coupled altermagnetic materials. Our framework reveals that the combined effects of magnon-induced spin flip and anisotropic spin relaxation of free electrons naturally give rise to a Néel-vector-dependent MR, manifesting as a $\pi$-periodic anisotropy that closely resembles its antiferromagnetic counterpart. Furthermore, we demonstrate that the interplay between the SHE and the SSE breaks this $\pi$ periodicity and induces a $2\pi$-periodic MR response. This transition originates from a linear dependence of the SHE-SSE coupling term on the Néel vector, enabling electrical distinction between binary Néel vector states.  Importantly, the resulting MR is not only highly tunable but also significantly amplified due to the potentially large spin conversion efficiency of the SSE ($\theta_{\text{SS}} \gg \theta_{\text{SH}}$), offering a pronounced contrast between high and low resistance states. This makes spin-orbit-coupled altermagnets ideal platforms for realizing electrically readable, high-efficiency altermagnetic memory devices. Beyond deepening the fundamental understanding of MR in altermagnetic systems, our work opens promising avenues for next-generation high-performance spintronic technologies based on altermagnetism.

\textit{Acknowledgment --} This work is supported by National Key R$\&$D Program of China (Grant No. 2020YFA0308800, No. 2022YFA1402600, and No. 2022YFA1403800), the National Natural Science Foundation of China (Grant No. 12234003, No. 12321004, No. 12274027, and No. 12374122),  and the Hong Kong Research Grants Council Grants (No. 16300523).

\onecolumngrid
\vspace{1em}
\begin{center}
	\textbf{\large End Matter}
\end{center}
\vspace{1em}
\twocolumngrid

\textit{Appendix A: Three Regimes for Spin Relaxation and MR Effect --} To qualitatively interpret Eqs.~\eqref{SL}–\eqref{vvh1}, we analyze three different magnetic regimes. (i) Nonmagnetic regime ($T \geq T_{\text{N}}$). In this regime, microscopic magnetic order is completely suppressed, leading to $\langle \text{S}_{\imath}^{\parallel} \rangle = 0$ and $\langle \text{S}_{\imath}^{\Vert} \text{S}_{\imath}^{\Vert} \rangle = \text{S}(\text{S}+1)/3$. As a result, both the Néel field from local moments and the spin-exchange field experienced by free electrons vanish. Spin relaxation becomes isotropic, i.e., $\tau^{-1}_{\Vert} = \tau^{-1}_{\perp} = \tau^{-1}_{0} + 2\Omega_1$, and no magnetoresistive effect is present.  (ii) Strongly magnetized (ferromagnetic) regime.  When the local spin moments are fully aligned (e.g., $\langle \text{S}_{\imath}^{\parallel} \rangle = -\text{S}$), spin-flip scattering of free electrons is completely suppressed ($\langle \text{S}_{\imath}^{\Vert} \text{S}_{\imath}^{\Vert} \rangle = \text{S}^2$). This leads to the maximum anisotropy in spin relaxation times: $\tau^{-1}_{\Vert} = \tau^{-1}_{0}$ and $\tau^{-1}_{\perp} = \tau^{-1}_{0} + 2\Omega_0$. Such anisotropy gives rise to a pronounced MR. However, the parallel alignment of local spins cancels the Néel field ($\mathcal{B}_{\text{N}} \propto \langle S_b^{\Vert} \rangle - \langle S_a^{\Vert} \rangle = 0$), thereby eliminating the SSE. (iii) Altermagnetic regime ($T < T_{\text{N}}$).  In this regime, the strong internal exchange enforces a compensated antiparallel alignment of local spins ($\langle \text{S}_{a}^{\parallel} \rangle = -\langle \text{S}_{b}^{\parallel} \rangle = -\text{S}$), suppressing spin-flip scattering ($\langle \text{S}_{\imath}^{\Vert} \text{S}_{\imath}^{\Vert} \rangle = \text{S}^2$). The Néel field reaches its maximum value,
$\mathcal{B}_{\text{N}} \propto \langle S_b^{\Vert} \rangle - \langle S_a^{\Vert} \rangle = 2\text{S}$, resulting in a maximal SSE with $\theta_{\text{SS}} = \theta^{0}_{\text{SS}}$. Furthermore, this regime also exhibits highly anisotropic spin relaxation times, $\tau^{-1}_{\Vert} = \tau^{-1}_{0}$ and $\tau^{-1}_{\perp} = \tau^{-1}_{0} + 2\Omega_0$,
thereby producing a strong anisotropic MR.  Importantly, in this altermagnetic regime, the spin-exchange field vanishes, thereby eliminating spin-precession-induced MR anisotropy. This stands in sharp contrast to regime (ii), where spin-precession MR invariably coexists with spin-relaxation-induced MR, even in the absence of an external magnetic field~\cite{zhang2024microscopic}. In the altermagnetic case, by contrast, we realize a pure spin-relaxation-driven MR anisotropy at zero magnetic field and low temperature ($T < T_{\text{N}}$), characterized by anisotropic spin diffusion lengths ($\ell_{\Vert} \neq \ell_{\perp}$) and the absence of Hanle spin precession.

\begin{figure}[t]
\begin{center}
\includegraphics[width=0.48\textwidth]{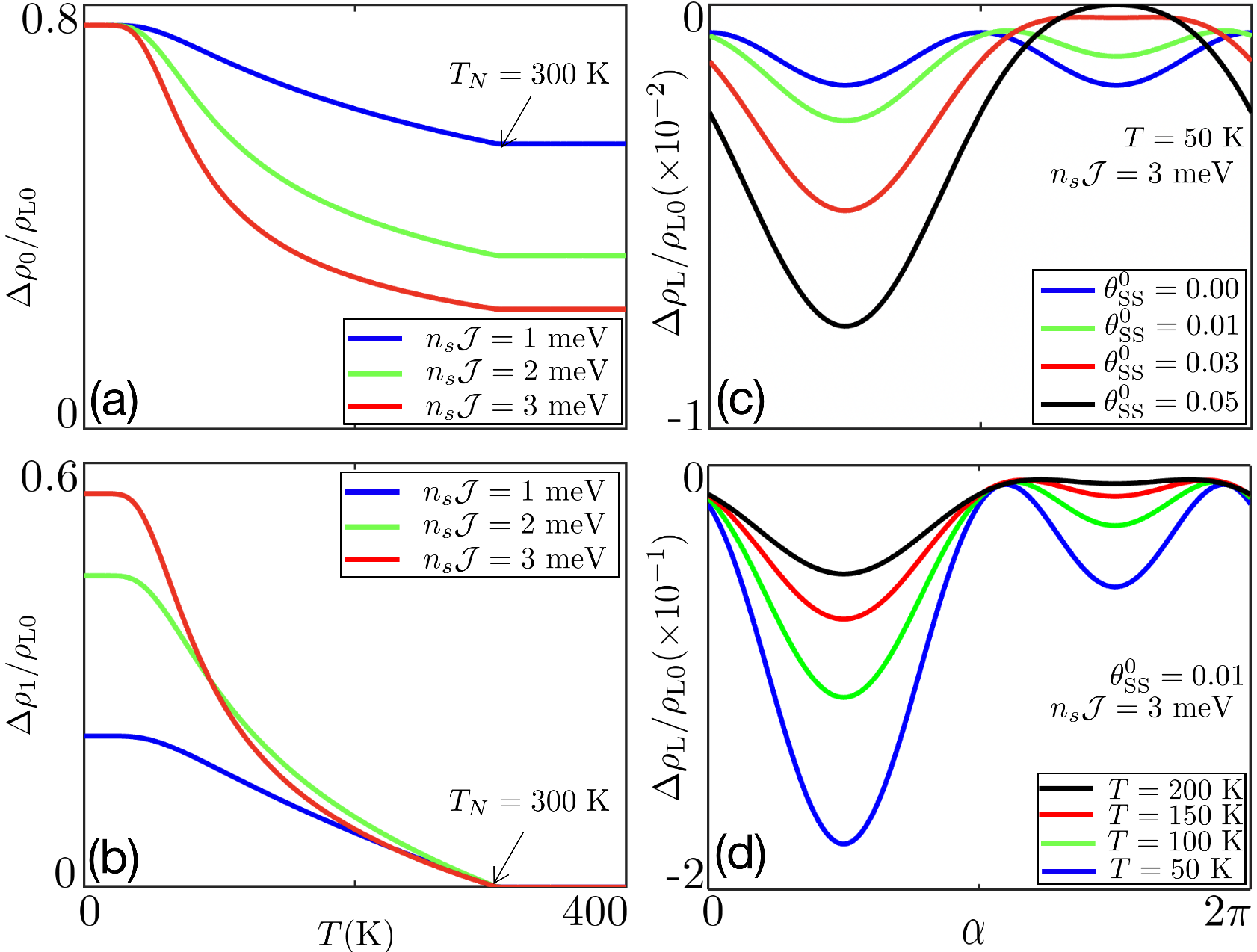} 
\end{center}
\caption{(a,b) The temperature ($T$) modulation of  $\Delta\rho _{0}$ and $\Delta\rho _{1}$ for several values of $n_s\mathcal{J}$. (c,d) Longitudinal resistivity as a function of the direction of the Néel vector ($n_y=\sin\alpha$) for different values of $\theta^0_{\text{SS}}$ and $T$. Here, $\text{S}=5/2$, $T_N=300$ K, and other parameters are same as Fig.~\ref{MMR}.}
\label{MMRRuO2}
\end{figure}

\textit{Appendix B: Two Charge-Spin Conversion Ratios --} Here, we derive the expressions for the charge-spin conversion ratios, $\theta_{\text{SS}}$ and $\theta_{\text{SH}}$, which originate from anisotropic spin-split energy bands and spin-orbit coupling, respectively. For clarity, we analyze these two contributions independently. The anisotropic spin-split energy spectrum of free electrons is given by $\epsilon_{\vec{k}\sigma}=\frac{\hbar^2k^2}{2m}+\sigma\lambda k^2\cos(2\theta_{\vec{k}}-2\Theta)/2$, where $\theta_{\vec{k}}$ is the azimuthal angle of momentum $\vec{k}$, and $\sigma^{\vec{n}}|\sigma\rangle = \sigma|\sigma\rangle$ with $\sigma = \pm 1$.  When an electric field is applied along the $x$-axis, the non-equilibrium deviation of the density matrix is approximated by $\delta\varrho_{\vec{k}\sigma}\simeq -e\tau_{} f'(\epsilon_{\vec{k}\sigma})v^x_{\vec{k}\sigma}E^x$, where $\tau$ is the momentum relaxation time due to randomly distributed impurities. The band velocities are expressed as $v^x_{\vec{k}\sigma}=v^{xx}_{k\sigma}\cos\theta_{\vec{k}}+v^{H}_{k\sigma}\sin\theta_{\vec{k}}$ and $v^z_{\vec{k}\sigma}=v^{zz}_{k\sigma}\sin\theta_{\vec{k}}+v^{H}_{k\sigma}\cos\theta_{\vec{k}}$ with spin-dependent components $v^{xx}_{k\sigma}=\frac{\hbar k}{m}+\sigma \lambda \hbar k\cos2\Theta$, $v^{zz}_{k\sigma}=\frac{\hbar k}{m}-\sigma\lambda \hbar k\cos2\Theta$, and $v^{H}_{k\sigma}=\sigma \lambda \hbar k\sin2\Theta$.  The charge and spin currents are defined as $\vec{J}_c = \vec{J}_\uparrow + \vec{J}_\downarrow$ and $\vec{J}_\text{SS} =  \vec{J}_\uparrow - \vec{J}_\downarrow$, respectively, where $J^{\nu}_{\sigma}=(e/V)\sum_{\vec{k}} v^{\nu}_{\vec{k}\sigma}\delta\varrho_{\vec{k}\sigma} $ ($\nu = {x, y, z}$) represents the contribution from spin-$\sigma$ carriers. Substituting the above expressions, we obtain the spin-splitter conversion ratio as
\begin{align} \label{svfvf}
    \theta_{\text{SS}}=\frac{J^{z}_\text{SS}}{J^x_c}\simeq\frac{2m\lambda \sin2\Theta}{1+m^2\lambda^2},
\end{align}
where we have neglected the $\theta_{\vec{k}}$-dependence of $f’(\epsilon_{\vec{k}\sigma})$ for brevity.

Next, we discuss the SHE arising from the extrinsic mechanism induced by impurity scattering. The band structure shown in Fig.~\ref{STORY}(a) does not include intrinsic spin-orbit coupling; instead, spin-orbit interaction is introduced through randomly distributed impurities located at positions $\vec{R}_j$, modeled by a localized potential $V_{\text{so}}(\vec{r})=U\sum_{j}\delta(\vec{r}-\vec{R}_j)$, where $U$ denotes the scattering strength.  The corresponding spin-orbit-modified scattering potential in momentum space takes the form $\sum_{\vec{k}\vec{k}'}[U^{o}_{\vec{k}\vec{k}'}-iU^{s}_{\vec{k}\vec{k}'}\vec{\sigma} \cdot (\vec{k} \times \vec{k}' )]$, where $U^{s}_{\vec{k}\vec{k}'}=l^2_{\text{so}}U\sum_je^{-i(\vec{k}-\vec{k}')\cdot\vec{R}_j}$ is the spin-orbit-dependent component, and $U^{o}_{\vec{k}\vec{k}'}=U\sum_je^{-i(\vec{k}-\vec{k}')\cdot\vec{R}_j}$ is an artificially introduced scalar, spin-independent term.  The spin-orbit interaction can be interpreted as generating an effective magnetic field along the spin-$\nu$ direction ($\nu = {x, y, z}$), giving rise to a spin current $\vec{J}^{\nu}_{\mathrm{SH}}$ with the flow direction, polarization direction ($\nu$), and the applied electric field $\vec{E}$ being mutually orthogonal: $\vec{J}^{\nu}_{\mathrm{SH}}=  \theta_{\text{SH}} \hat{\nu} \times \sigma_{\text{D}}\vec{E}$, as previously discussed in~\cite{chen2013theory}. Following the standard framework of quantum transport theory~\cite{sinova2015spin}, the spin Hall angle is obtained as
\begin{align} \label{svfv}
    \theta_{\text{SH}}=\frac{4\pi}{3}\nu_FU(k_Fl_{\text{so}})^2,
\end{align}
where $k_F$ is the Fermi wavevector and $\nu_F$ is the density of states at the Fermi level.

\textit{Appendix C: Numerical Results for Altermagnetic RuO$_2$ –} Finally, we present the MR behavior of RuO$_2$, one of the most debated altermagnets, which exhibits a Néel temperature near room temperature ($T_{\text{N}} \simeq 300$ K)~\cite{feng2024incommensurate,bai2022observation}. In our analysis, we consider (010)-oriented RuO$_2$ thin films, where the spin quantization axis lies along the $y$-direction, parallel to the polarization direction of the SHE-induced spin current. The MR effect in RuO$_2$ arises from a two-step charge–spin–charge conversion process involving both the SHE and SSE, which are known to coexist in this material~\cite{liao2024separation,feng2024incommensurate}. The numerical results, presented in Fig.~\ref{MMRRuO2}, closely resemble those obtained for CrSb in Fig.~\ref{MMR}, indicating the generality of our theoretical framework.  While earlier studies have questioned the existence of an altermagnetic ground state in RuO$_2$~\cite{hiraishi2024nonmagnetic,liu2024absence,kessler2024absence}, recent experiments on RuO$_2$ thin films have provided direct evidence of its altermagnetic character, notably through the observation of spin-dependent tunneling MR~\cite{noh2025tunneling}. The distinct anisotropic MR signatures predicted in our work offer a direct and robust method to resolve the ongoing debate regarding the altermagnetic nature of RuO$_2$.


\begin{thebibliography}{47}%
\makeatletter
\providecommand \@ifxundefined [1]{%
 \@ifx{#1\undefined}
}%
\providecommand \@ifnum [1]{%
 \ifnum #1\expandafter \@firstoftwo
 \else \expandafter \@secondoftwo
 \fi
}%
\providecommand \@ifx [1]{%
 \ifx #1\expandafter \@firstoftwo
 \else \expandafter \@secondoftwo
 \fi
}%
\providecommand \natexlab [1]{#1}%
\providecommand \enquote  [1]{``#1''}%
\providecommand \bibnamefont  [1]{#1}%
\providecommand \bibfnamefont [1]{#1}%
\providecommand \citenamefont [1]{#1}%
\providecommand \href@noop [0]{\@secondoftwo}%
\providecommand \href [0]{\begingroup \@sanitize@url \@href}%
\providecommand \@href[1]{\@@startlink{#1}\@@href}%
\providecommand \@@href[1]{\endgroup#1\@@endlink}%
\providecommand \@sanitize@url [0]{\catcode `\\12\catcode `\$12\catcode
  `\&12\catcode `\#12\catcode `\^12\catcode `\_12\catcode `\%12\relax}%
\providecommand \@@startlink[1]{}%
\providecommand \@@endlink[0]{}%
\providecommand \url  [0]{\begingroup\@sanitize@url \@url }%
\providecommand \@url [1]{\endgroup\@href {#1}{\urlprefix }}%
\providecommand \urlprefix  [0]{URL }%
\providecommand \Eprint [0]{\href }%
\providecommand \doibase [0]{https://doi.org/}%
\providecommand \selectlanguage [0]{\@gobble}%
\providecommand \bibinfo  [0]{\@secondoftwo}%
\providecommand \bibfield  [0]{\@secondoftwo}%
\providecommand \translation [1]{[#1]}%
\providecommand \BibitemOpen [0]{}%
\providecommand \bibitemStop [0]{}%
\providecommand \bibitemNoStop [0]{.\EOS\space}%
\providecommand \EOS [0]{\spacefactor3000\relax}%
\providecommand \BibitemShut  [1]{\csname bibitem#1\endcsname}%
\let\auto@bib@innerbib\@empty
%</preamble>
\bibitem [{\citenamefont {\ifmmode~\check{S}\else \v{S}\fi{}mejkal}\ \emph
  {et~al.}(2022{\natexlab{a}})\citenamefont {\ifmmode~\check{S}\else
  \v{S}\fi{}mejkal}, \citenamefont {Sinova},\ and\ \citenamefont
  {Jungwirth}}]{libor2022beyond}%
  \BibitemOpen
  \bibfield  {author} {\bibinfo {author} {\bibfnamefont {L.}~\bibnamefont
  {\ifmmode~\check{S}\else \v{S}\fi{}mejkal}}, \bibinfo {author} {\bibfnamefont
  {J.}~\bibnamefont {Sinova}},\ and\ \bibinfo {author} {\bibfnamefont
  {T.}~\bibnamefont {Jungwirth}},\ }\bibfield  {title} {\bibinfo {title}
  {Beyond conventional ferromagnetism and antiferromagnetism: A phase with
  nonrelativistic spin and crystal rotation symmetry},\ }\href
  {https://doi.org/10.1103/PhysRevX.12.031042} {\bibfield  {journal} {\bibinfo
  {journal} {Phys. Rev. X}\ }\textbf {\bibinfo {volume} {12}},\ \bibinfo
  {pages} {031042} (\bibinfo {year} {2022}{\natexlab{a}})}\BibitemShut
  {NoStop}%
\bibitem [{\citenamefont {\ifmmode~\check{S}\else \v{S}\fi{}mejkal}\ \emph
  {et~al.}(2022{\natexlab{b}})\citenamefont {\ifmmode~\check{S}\else
  \v{S}\fi{}mejkal}, \citenamefont {Sinova},\ and\ \citenamefont
  {Jungwirth}}]{libor2024emerging}%
  \BibitemOpen
  \bibfield  {author} {\bibinfo {author} {\bibfnamefont {L.}~\bibnamefont
  {\ifmmode~\check{S}\else \v{S}\fi{}mejkal}}, \bibinfo {author} {\bibfnamefont
  {J.}~\bibnamefont {Sinova}},\ and\ \bibinfo {author} {\bibfnamefont
  {T.}~\bibnamefont {Jungwirth}},\ }\bibfield  {title} {\bibinfo {title}
  {Emerging research landscape of altermagnetism},\ }\href
  {https://doi.org/10.1103/PhysRevX.12.040501} {\bibfield  {journal} {\bibinfo
  {journal} {Phys. Rev. X}\ }\textbf {\bibinfo {volume} {12}},\ \bibinfo
  {pages} {040501} (\bibinfo {year} {2022}{\natexlab{b}})}\BibitemShut
  {NoStop}%
\bibitem [{\citenamefont {Bai}\ \emph {et~al.}(2024)\citenamefont {Bai},
  \citenamefont {Feng}, \citenamefont {Liu}, \citenamefont {{\v S}mejkal},
  \citenamefont {Mokrousov},\ and\ \citenamefont
  {Yao}}]{bai2024altermagnetism}%
  \BibitemOpen
  \bibfield  {author} {\bibinfo {author} {\bibfnamefont {L.}~\bibnamefont
  {Bai}}, \bibinfo {author} {\bibfnamefont {W.}~\bibnamefont {Feng}}, \bibinfo
  {author} {\bibfnamefont {S.}~\bibnamefont {Liu}}, \bibinfo {author}
  {\bibfnamefont {L.}~\bibnamefont {{\v S}mejkal}}, \bibinfo {author}
  {\bibfnamefont {Y.}~\bibnamefont {Mokrousov}},\ and\ \bibinfo {author}
  {\bibfnamefont {Y.}~\bibnamefont {Yao}},\ }\bibfield  {title} {\bibinfo
  {title} {Altermagnetism: Exploring new frontiers in magnetism and
  spintronics},\ }\href
  {https://doi.org/https://doi.org/10.1002/adfm.202409327} {\bibfield
  {journal} {\bibinfo  {journal} {Adv. Funct. Mater.}\ }\textbf {\bibinfo
  {volume} {34}},\ \bibinfo {pages} {2409327} (\bibinfo {year}
  {2024})}\BibitemShut {NoStop}%
\bibitem [{\citenamefont {Fedchenko}\ \emph {et~al.}(2024)\citenamefont
  {Fedchenko}, \citenamefont {Min{\'a}r}, \citenamefont {Akashdeep},
  \citenamefont {D'Souza}, \citenamefont {Vasilyev}, \citenamefont {Tkach},
  \citenamefont {Odenbreit}, \citenamefont {Nguyen}, \citenamefont
  {Kutnyakhov}, \citenamefont {Wind}, \citenamefont {Wenthaus}, \citenamefont
  {Scholz}, \citenamefont {Rossnagel}, \citenamefont {Hoesch}, \citenamefont
  {Aeschlimann}, \citenamefont {Stadtm{\"u}ller}, \citenamefont {Kl{\"a}ui},
  \citenamefont {Sch{\"o}nhense}, \citenamefont {Jungwirth}, \citenamefont
  {Hellenes}, \citenamefont {Jakob}, \citenamefont {{\v S}mejkal},
  \citenamefont {Sinova},\ and\ \citenamefont {Elmers}}]{Fedchenko2024}%
  \BibitemOpen
  \bibfield  {author} {\bibinfo {author} {\bibfnamefont {O.}~\bibnamefont
  {Fedchenko}}, \bibinfo {author} {\bibfnamefont {J.}~\bibnamefont
  {Min{\'a}r}}, \bibinfo {author} {\bibfnamefont {A.}~\bibnamefont
  {Akashdeep}}, \bibinfo {author} {\bibfnamefont {S.~W.}\ \bibnamefont
  {D'Souza}}, \bibinfo {author} {\bibfnamefont {D.}~\bibnamefont {Vasilyev}},
  \bibinfo {author} {\bibfnamefont {O.}~\bibnamefont {Tkach}}, \bibinfo
  {author} {\bibfnamefont {L.}~\bibnamefont {Odenbreit}}, \bibinfo {author}
  {\bibfnamefont {Q.}~\bibnamefont {Nguyen}}, \bibinfo {author} {\bibfnamefont
  {D.}~\bibnamefont {Kutnyakhov}}, \bibinfo {author} {\bibfnamefont
  {N.}~\bibnamefont {Wind}}, \bibinfo {author} {\bibfnamefont {L.}~\bibnamefont
  {Wenthaus}}, \bibinfo {author} {\bibfnamefont {M.}~\bibnamefont {Scholz}},
  \bibinfo {author} {\bibfnamefont {K.}~\bibnamefont {Rossnagel}}, \bibinfo
  {author} {\bibfnamefont {M.}~\bibnamefont {Hoesch}}, \bibinfo {author}
  {\bibfnamefont {M.}~\bibnamefont {Aeschlimann}}, \bibinfo {author}
  {\bibfnamefont {B.}~\bibnamefont {Stadtm{\"u}ller}}, \bibinfo {author}
  {\bibfnamefont {M.}~\bibnamefont {Kl{\"a}ui}}, \bibinfo {author}
  {\bibfnamefont {G.}~\bibnamefont {Sch{\"o}nhense}}, \bibinfo {author}
  {\bibfnamefont {T.}~\bibnamefont {Jungwirth}}, \bibinfo {author}
  {\bibfnamefont {A.~B.}\ \bibnamefont {Hellenes}}, \bibinfo {author}
  {\bibfnamefont {G.}~\bibnamefont {Jakob}}, \bibinfo {author} {\bibfnamefont
  {L.}~\bibnamefont {{\v S}mejkal}}, \bibinfo {author} {\bibfnamefont
  {J.}~\bibnamefont {Sinova}},\ and\ \bibinfo {author} {\bibfnamefont {H.-J.}\
  \bibnamefont {Elmers}},\ }\bibfield  {title} {\bibinfo {title} {Observation
  of time-reversal symmetry breaking in the band structure of altermagnetic
  \text{RuO}$_2$},\ }\href {https://doi.org/10.1126/sciadv.adj4883} {\bibfield
  {journal} {\bibinfo  {journal} {Sci. Adv.}\ }\textbf {\bibinfo {volume}
  {10}},\ \bibinfo {pages} {eadj4883} (\bibinfo {year} {2024})}\BibitemShut
  {NoStop}%
\bibitem [{\citenamefont {Reimers}\ \emph {et~al.}(2024)\citenamefont
  {Reimers}, \citenamefont {Odenbreit}, \citenamefont {{\v S}mejkal},
  \citenamefont {Strocov}, \citenamefont {Constantinou}, \citenamefont
  {Hellenes}, \citenamefont {Jaeschke~Ubiergo}, \citenamefont {Campos},
  \citenamefont {Bharadwaj}, \citenamefont {Chakraborty}, \citenamefont
  {Denneulin}, \citenamefont {Shi}, \citenamefont {Dunin-Borkowski},
  \citenamefont {Das}, \citenamefont {Kl{\"a}ui}, \citenamefont {Sinova},\ and\
  \citenamefont {Jourdan}}]{Reimers2024}%
  \BibitemOpen
  \bibfield  {author} {\bibinfo {author} {\bibfnamefont {S.}~\bibnamefont
  {Reimers}}, \bibinfo {author} {\bibfnamefont {L.}~\bibnamefont {Odenbreit}},
  \bibinfo {author} {\bibfnamefont {L.}~\bibnamefont {{\v S}mejkal}}, \bibinfo
  {author} {\bibfnamefont {V.~N.}\ \bibnamefont {Strocov}}, \bibinfo {author}
  {\bibfnamefont {P.}~\bibnamefont {Constantinou}}, \bibinfo {author}
  {\bibfnamefont {A.~B.}\ \bibnamefont {Hellenes}}, \bibinfo {author}
  {\bibfnamefont {R.}~\bibnamefont {Jaeschke~Ubiergo}}, \bibinfo {author}
  {\bibfnamefont {W.~H.}\ \bibnamefont {Campos}}, \bibinfo {author}
  {\bibfnamefont {V.~K.}\ \bibnamefont {Bharadwaj}}, \bibinfo {author}
  {\bibfnamefont {A.}~\bibnamefont {Chakraborty}}, \bibinfo {author}
  {\bibfnamefont {T.}~\bibnamefont {Denneulin}}, \bibinfo {author}
  {\bibfnamefont {W.}~\bibnamefont {Shi}}, \bibinfo {author} {\bibfnamefont
  {R.~E.}\ \bibnamefont {Dunin-Borkowski}}, \bibinfo {author} {\bibfnamefont
  {S.}~\bibnamefont {Das}}, \bibinfo {author} {\bibfnamefont {M.}~\bibnamefont
  {Kl{\"a}ui}}, \bibinfo {author} {\bibfnamefont {J.}~\bibnamefont {Sinova}},\
  and\ \bibinfo {author} {\bibfnamefont {M.}~\bibnamefont {Jourdan}},\
  }\bibfield  {title} {\bibinfo {title} {Direct observation of altermagnetic
  band splitting in \text{CrSb} thin films},\ }\href
  {https://doi.org/10.1038/s41467-024-46476-5} {\bibfield  {journal} {\bibinfo
  {journal} {Nat. Commun.}\ }\textbf {\bibinfo {volume} {15}},\ \bibinfo
  {pages} {2116} (\bibinfo {year} {2024})}\BibitemShut {NoStop}%
\bibitem [{\citenamefont {Krempask{\'y}}\ \emph {et~al.}(2024)\citenamefont
  {Krempask{\'y}}, \citenamefont {{\v S}mejkal}, \citenamefont {D'Souza},
  \citenamefont {Hajlaoui}, \citenamefont {Springholz}, \citenamefont
  {Uhl{\'\i}{\v r}ov{\'a}}, \citenamefont {Alarab}, \citenamefont
  {Constantinou}, \citenamefont {Strocov}, \citenamefont {Usanov},
  \citenamefont {Pudelko}, \citenamefont {Gonz{\'a}lez-Hern{\'a}ndez},
  \citenamefont {Birk~Hellenes}, \citenamefont {Jansa}, \citenamefont
  {Reichlov{\'a}}, \citenamefont {{\v S}ob{\'a}{\v n}}, \citenamefont
  {Gonzalez~Betancourt}, \citenamefont {Wadley}, \citenamefont {Sinova},
  \citenamefont {Kriegner}, \citenamefont {Min{\'a}r}, \citenamefont {Dil},\
  and\ \citenamefont {Jungwirth}}]{Krempasky2024}%
  \BibitemOpen
  \bibfield  {author} {\bibinfo {author} {\bibfnamefont {J.}~\bibnamefont
  {Krempask{\'y}}}, \bibinfo {author} {\bibfnamefont {L.}~\bibnamefont {{\v
  S}mejkal}}, \bibinfo {author} {\bibfnamefont {S.~W.}\ \bibnamefont
  {D'Souza}}, \bibinfo {author} {\bibfnamefont {M.}~\bibnamefont {Hajlaoui}},
  \bibinfo {author} {\bibfnamefont {G.}~\bibnamefont {Springholz}}, \bibinfo
  {author} {\bibfnamefont {K.}~\bibnamefont {Uhl{\'\i}{\v r}ov{\'a}}}, \bibinfo
  {author} {\bibfnamefont {F.}~\bibnamefont {Alarab}}, \bibinfo {author}
  {\bibfnamefont {P.~C.}\ \bibnamefont {Constantinou}}, \bibinfo {author}
  {\bibfnamefont {V.}~\bibnamefont {Strocov}}, \bibinfo {author} {\bibfnamefont
  {D.}~\bibnamefont {Usanov}}, \bibinfo {author} {\bibfnamefont {W.~R.}\
  \bibnamefont {Pudelko}}, \bibinfo {author} {\bibfnamefont {R.}~\bibnamefont
  {Gonz{\'a}lez-Hern{\'a}ndez}}, \bibinfo {author} {\bibfnamefont
  {A.}~\bibnamefont {Birk~Hellenes}}, \bibinfo {author} {\bibfnamefont
  {Z.}~\bibnamefont {Jansa}}, \bibinfo {author} {\bibfnamefont
  {H.}~\bibnamefont {Reichlov{\'a}}}, \bibinfo {author} {\bibfnamefont
  {Z.}~\bibnamefont {{\v S}ob{\'a}{\v n}}}, \bibinfo {author} {\bibfnamefont
  {R.~D.}\ \bibnamefont {Gonzalez~Betancourt}}, \bibinfo {author}
  {\bibfnamefont {P.}~\bibnamefont {Wadley}}, \bibinfo {author} {\bibfnamefont
  {J.}~\bibnamefont {Sinova}}, \bibinfo {author} {\bibfnamefont
  {D.}~\bibnamefont {Kriegner}}, \bibinfo {author} {\bibfnamefont
  {J.}~\bibnamefont {Min{\'a}r}}, \bibinfo {author} {\bibfnamefont {J.~H.}\
  \bibnamefont {Dil}},\ and\ \bibinfo {author} {\bibfnamefont {T.}~\bibnamefont
  {Jungwirth}},\ }\bibfield  {title} {\bibinfo {title} {Altermagnetic lifting
  of kramers spin degeneracy},\ }\href
  {https://doi.org/10.1038/s41586-023-06907-7} {\bibfield  {journal} {\bibinfo
  {journal} {Nature}\ }\textbf {\bibinfo {volume} {626}},\ \bibinfo {pages}
  {517} (\bibinfo {year} {2024})}\BibitemShut {NoStop}%
\bibitem [{\citenamefont {Gonz\'alez-Hern\'andez}\ \emph
  {et~al.}(2021)\citenamefont {Gonz\'alez-Hern\'andez}, \citenamefont
  {\ifmmode~\check{S}\else \v{S}\fi{}mejkal}, \citenamefont {V\'yborn\'y},
  \citenamefont {Yahagi}, \citenamefont {Sinova}, \citenamefont {Jungwirth},\
  and\ \citenamefont {\ifmmode~\check{Z}\else
  \v{Z}\fi{}elezn\'y}}]{gonzalez2021efficient}%
  \BibitemOpen
  \bibfield  {author} {\bibinfo {author} {\bibfnamefont {R.}~\bibnamefont
  {Gonz\'alez-Hern\'andez}}, \bibinfo {author} {\bibfnamefont {L.}~\bibnamefont
  {\ifmmode~\check{S}\else \v{S}\fi{}mejkal}}, \bibinfo {author} {\bibfnamefont
  {K.}~\bibnamefont {V\'yborn\'y}}, \bibinfo {author} {\bibfnamefont
  {Y.}~\bibnamefont {Yahagi}}, \bibinfo {author} {\bibfnamefont
  {J.}~\bibnamefont {Sinova}}, \bibinfo {author} {\bibfnamefont {T.~c.~v.}\
  \bibnamefont {Jungwirth}},\ and\ \bibinfo {author} {\bibfnamefont
  {J.}~\bibnamefont {\ifmmode~\check{Z}\else \v{Z}\fi{}elezn\'y}},\ }\bibfield
  {title} {\bibinfo {title} {Efficient electrical spin splitter based on
  nonrelativistic collinear antiferromagnetism},\ }\href
  {https://doi.org/10.1103/PhysRevLett.126.127701} {\bibfield  {journal}
  {\bibinfo  {journal} {Phys. Rev. Lett.}\ }\textbf {\bibinfo {volume} {126}},\
  \bibinfo {pages} {127701} (\bibinfo {year} {2021})}\BibitemShut {NoStop}%
\bibitem [{\citenamefont {Bai}\ \emph {et~al.}(2022)\citenamefont {Bai},
  \citenamefont {Han}, \citenamefont {Feng}, \citenamefont {Zhou},
  \citenamefont {Su}, \citenamefont {Wang}, \citenamefont {Liao}, \citenamefont
  {Zhu}, \citenamefont {Chen}, \citenamefont {Pan}, \citenamefont {Fan},\ and\
  \citenamefont {Song}}]{bai2022observation}%
  \BibitemOpen
  \bibfield  {author} {\bibinfo {author} {\bibfnamefont {H.}~\bibnamefont
  {Bai}}, \bibinfo {author} {\bibfnamefont {L.}~\bibnamefont {Han}}, \bibinfo
  {author} {\bibfnamefont {X.~Y.}\ \bibnamefont {Feng}}, \bibinfo {author}
  {\bibfnamefont {Y.~J.}\ \bibnamefont {Zhou}}, \bibinfo {author}
  {\bibfnamefont {R.~X.}\ \bibnamefont {Su}}, \bibinfo {author} {\bibfnamefont
  {Q.}~\bibnamefont {Wang}}, \bibinfo {author} {\bibfnamefont {L.~Y.}\
  \bibnamefont {Liao}}, \bibinfo {author} {\bibfnamefont {W.~X.}\ \bibnamefont
  {Zhu}}, \bibinfo {author} {\bibfnamefont {X.~Z.}\ \bibnamefont {Chen}},
  \bibinfo {author} {\bibfnamefont {F.}~\bibnamefont {Pan}}, \bibinfo {author}
  {\bibfnamefont {X.~L.}\ \bibnamefont {Fan}},\ and\ \bibinfo {author}
  {\bibfnamefont {C.}~\bibnamefont {Song}},\ }\bibfield  {title} {\bibinfo
  {title} {Observation of spin splitting torque in a collinear antiferromagnet
  \text{RuO}$_{2}$},\ }\href {https://doi.org/10.1103/PhysRevLett.128.197202}
  {\bibfield  {journal} {\bibinfo  {journal} {Phys. Rev. Lett.}\ }\textbf
  {\bibinfo {volume} {128}},\ \bibinfo {pages} {197202} (\bibinfo {year}
  {2022})}\BibitemShut {NoStop}%
\bibitem [{\citenamefont {{\v S}mejkal}\ \emph {et~al.}(2020)\citenamefont {{\v
  S}mejkal}, \citenamefont {Gonz{\'a}lez-Hern{\'a}ndez}, \citenamefont
  {Jungwirth},\ and\ \citenamefont {Sinova}}]{libor2020AHE}%
  \BibitemOpen
  \bibfield  {author} {\bibinfo {author} {\bibfnamefont {L.}~\bibnamefont {{\v
  S}mejkal}}, \bibinfo {author} {\bibfnamefont {R.}~\bibnamefont
  {Gonz{\'a}lez-Hern{\'a}ndez}}, \bibinfo {author} {\bibfnamefont
  {T.}~\bibnamefont {Jungwirth}},\ and\ \bibinfo {author} {\bibfnamefont
  {J.}~\bibnamefont {Sinova}},\ }\bibfield  {title} {\bibinfo {title} {Crystal
  time-reversal symmetry breaking and spontaneous \text{H}all effect in
  collinear antiferromagnets},\ }\href {https://doi.org/10.1126/sciadv.aaz8809}
  {\bibfield  {journal} {\bibinfo  {journal} {Sci. Adv.}\ }\textbf {\bibinfo
  {volume} {6}},\ \bibinfo {pages} {eaaz8809} (\bibinfo {year}
  {2020})}\BibitemShut {NoStop}%
\bibitem [{\citenamefont {Feng}\ \emph {et~al.}(2022)\citenamefont {Feng},
  \citenamefont {Zhou}, \citenamefont {{\v{S}}mejkal}, \citenamefont {Wu},
  \citenamefont {Zhu}, \citenamefont {Guo}, \citenamefont
  {Gonz{\'a}lez-Hern{\'a}ndez}, \citenamefont {Wang}, \citenamefont {Yan},
  \citenamefont {Qin} \emph {et~al.}}]{feng2022anomalous}%
  \BibitemOpen
  \bibfield  {author} {\bibinfo {author} {\bibfnamefont {Z.}~\bibnamefont
  {Feng}}, \bibinfo {author} {\bibfnamefont {X.}~\bibnamefont {Zhou}}, \bibinfo
  {author} {\bibfnamefont {L.}~\bibnamefont {{\v{S}}mejkal}}, \bibinfo {author}
  {\bibfnamefont {L.}~\bibnamefont {Wu}}, \bibinfo {author} {\bibfnamefont
  {Z.}~\bibnamefont {Zhu}}, \bibinfo {author} {\bibfnamefont {H.}~\bibnamefont
  {Guo}}, \bibinfo {author} {\bibfnamefont {R.}~\bibnamefont
  {Gonz{\'a}lez-Hern{\'a}ndez}}, \bibinfo {author} {\bibfnamefont
  {X.}~\bibnamefont {Wang}}, \bibinfo {author} {\bibfnamefont {H.}~\bibnamefont
  {Yan}}, \bibinfo {author} {\bibfnamefont {P.}~\bibnamefont {Qin}}, \emph
  {et~al.},\ }\bibfield  {title} {\bibinfo {title} {An anomalous \text{H}all
  effect in altermagnetic ruthenium dioxide},\ }\href
  {https://doi.org/https://doi.org/10.1038/s41928-022-00866-z} {\bibfield
  {journal} {\bibinfo  {journal} {Nat. Electron.}\ }\textbf {\bibinfo {volume}
  {5}},\ \bibinfo {pages} {735} (\bibinfo {year} {2022})}\BibitemShut {NoStop}%
\bibitem [{\citenamefont {Zhou}\ \emph {et~al.}(2021)\citenamefont {Zhou},
  \citenamefont {Feng}, \citenamefont {Yang}, \citenamefont {Guo},\ and\
  \citenamefont {Yao}}]{zhou2021crystal}%
  \BibitemOpen
  \bibfield  {author} {\bibinfo {author} {\bibfnamefont {X.}~\bibnamefont
  {Zhou}}, \bibinfo {author} {\bibfnamefont {W.}~\bibnamefont {Feng}}, \bibinfo
  {author} {\bibfnamefont {X.}~\bibnamefont {Yang}}, \bibinfo {author}
  {\bibfnamefont {G.-Y.}\ \bibnamefont {Guo}},\ and\ \bibinfo {author}
  {\bibfnamefont {Y.}~\bibnamefont {Yao}},\ }\bibfield  {title} {\bibinfo
  {title} {Crystal chirality magneto-optical effects in collinear
  antiferromagnets},\ }\href {https://doi.org/10.1103/PhysRevB.104.024401}
  {\bibfield  {journal} {\bibinfo  {journal} {Phys. Rev. B}\ }\textbf {\bibinfo
  {volume} {104}},\ \bibinfo {pages} {024401} (\bibinfo {year}
  {2021})}\BibitemShut {NoStop}%
\bibitem [{\citenamefont {Hariki}\ \emph {et~al.}(2024)\citenamefont {Hariki},
  \citenamefont {Dal~Din}, \citenamefont {Amin}, \citenamefont {Yamaguchi},
  \citenamefont {Badura}, \citenamefont {Kriegner}, \citenamefont {Edmonds},
  \citenamefont {Campion}, \citenamefont {Wadley}, \citenamefont {Backes},
  \citenamefont {Veiga}, \citenamefont {Dhesi}, \citenamefont {Springholz},
  \citenamefont {\ifmmode~\check{S}\else \v{S}\fi{}mejkal}, \citenamefont
  {V\'yborn\'y}, \citenamefont {Jungwirth},\ and\ \citenamefont
  {Kune\ifmmode~\check{s}\else \v{s}\fi{}}}]{Hariki2024}%
  \BibitemOpen
  \bibfield  {author} {\bibinfo {author} {\bibfnamefont {A.}~\bibnamefont
  {Hariki}}, \bibinfo {author} {\bibfnamefont {A.}~\bibnamefont {Dal~Din}},
  \bibinfo {author} {\bibfnamefont {O.~J.}\ \bibnamefont {Amin}}, \bibinfo
  {author} {\bibfnamefont {T.}~\bibnamefont {Yamaguchi}}, \bibinfo {author}
  {\bibfnamefont {A.}~\bibnamefont {Badura}}, \bibinfo {author} {\bibfnamefont
  {D.}~\bibnamefont {Kriegner}}, \bibinfo {author} {\bibfnamefont {K.~W.}\
  \bibnamefont {Edmonds}}, \bibinfo {author} {\bibfnamefont {R.~P.}\
  \bibnamefont {Campion}}, \bibinfo {author} {\bibfnamefont {P.}~\bibnamefont
  {Wadley}}, \bibinfo {author} {\bibfnamefont {D.}~\bibnamefont {Backes}},
  \bibinfo {author} {\bibfnamefont {L.~S.~I.}\ \bibnamefont {Veiga}}, \bibinfo
  {author} {\bibfnamefont {S.~S.}\ \bibnamefont {Dhesi}}, \bibinfo {author}
  {\bibfnamefont {G.}~\bibnamefont {Springholz}}, \bibinfo {author}
  {\bibfnamefont {L.}~\bibnamefont {\ifmmode~\check{S}\else \v{S}\fi{}mejkal}},
  \bibinfo {author} {\bibfnamefont {K.}~\bibnamefont {V\'yborn\'y}}, \bibinfo
  {author} {\bibfnamefont {T.}~\bibnamefont {Jungwirth}},\ and\ \bibinfo
  {author} {\bibfnamefont {J.}~\bibnamefont {Kune\ifmmode~\check{s}\else
  \v{s}\fi{}}},\ }\bibfield  {title} {\bibinfo {title} {X-ray magnetic circular
  dichroism in altermagnetic $\ensuremath{\alpha}$-mnte},\ }\href
  {https://doi.org/10.1103/PhysRevLett.132.176701} {\bibfield  {journal}
  {\bibinfo  {journal} {Phys. Rev. Lett.}\ }\textbf {\bibinfo {volume} {132}},\
  \bibinfo {pages} {176701} (\bibinfo {year} {2024})}\BibitemShut {NoStop}%
\bibitem [{\citenamefont {Gray}\ \emph {et~al.}(2024)\citenamefont {Gray},
  \citenamefont {Deng}, \citenamefont {Tian}, \citenamefont {Chilcote},
  \citenamefont {Dodge}, \citenamefont {Brahlek},\ and\ \citenamefont
  {Wu}}]{Gray2024}%
  \BibitemOpen
  \bibfield  {author} {\bibinfo {author} {\bibfnamefont {I.}~\bibnamefont
  {Gray}}, \bibinfo {author} {\bibfnamefont {Q.}~\bibnamefont {Deng}}, \bibinfo
  {author} {\bibfnamefont {Q.}~\bibnamefont {Tian}}, \bibinfo {author}
  {\bibfnamefont {M.}~\bibnamefont {Chilcote}}, \bibinfo {author}
  {\bibfnamefont {J.~S.}\ \bibnamefont {Dodge}}, \bibinfo {author}
  {\bibfnamefont {M.}~\bibnamefont {Brahlek}},\ and\ \bibinfo {author}
  {\bibfnamefont {L.}~\bibnamefont {Wu}},\ }\bibfield  {title} {\bibinfo
  {title} {Time-resolved magneto-optical effects in the altermagnet candidate
  \text{MnTe}},\ }\href {https://arxiv.org/pdf/2404.05020.pdf} {\bibfield
  {journal} {\bibinfo  {journal} {arXiv}\ } (\bibinfo {year} {2024})},\ \Eprint
  {https://arxiv.org/abs/2404.05020} {2404.05020} \BibitemShut {NoStop}%
\bibitem [{\citenamefont {Han}\ \emph {et~al.}(2024)\citenamefont {Han},
  \citenamefont {Fu}, \citenamefont {Peng}, \citenamefont {Cheng},
  \citenamefont {Dai}, \citenamefont {Liu}, \citenamefont {Li}, \citenamefont
  {Zhang}, \citenamefont {Zhu}, \citenamefont {Bai} \emph
  {et~al.}}]{han2024electrical}%
  \BibitemOpen
  \bibfield  {author} {\bibinfo {author} {\bibfnamefont {L.}~\bibnamefont
  {Han}}, \bibinfo {author} {\bibfnamefont {X.}~\bibnamefont {Fu}}, \bibinfo
  {author} {\bibfnamefont {R.}~\bibnamefont {Peng}}, \bibinfo {author}
  {\bibfnamefont {X.}~\bibnamefont {Cheng}}, \bibinfo {author} {\bibfnamefont
  {J.}~\bibnamefont {Dai}}, \bibinfo {author} {\bibfnamefont {L.}~\bibnamefont
  {Liu}}, \bibinfo {author} {\bibfnamefont {Y.}~\bibnamefont {Li}}, \bibinfo
  {author} {\bibfnamefont {Y.}~\bibnamefont {Zhang}}, \bibinfo {author}
  {\bibfnamefont {W.}~\bibnamefont {Zhu}}, \bibinfo {author} {\bibfnamefont
  {H.}~\bibnamefont {Bai}}, \emph {et~al.},\ }\bibfield  {title} {\bibinfo
  {title} {Electrical 180$\,^{\circ}$ switching of n{\'e}el vector in
  spin-splitting antiferromagnet},\ }\href
  {https://doi.org/10.1126/sciadv.adn0479} {\bibfield  {journal} {\bibinfo
  {journal} {Sci. Adv.}\ }\textbf {\bibinfo {volume} {10}},\ \bibinfo {pages}
  {eadn0479} (\bibinfo {year} {2024})}\BibitemShut {NoStop}%
\bibitem [{\citenamefont {\ifmmode~\check{S}\else \v{S}\fi{}mejkal}\ \emph
  {et~al.}(2022{\natexlab{c}})\citenamefont {\ifmmode~\check{S}\else
  \v{S}\fi{}mejkal}, \citenamefont {Hellenes}, \citenamefont
  {Gonz\'alez-Hern\'andez}, \citenamefont {Sinova},\ and\ \citenamefont
  {Jungwirth}}]{libor2022giant}%
  \BibitemOpen
  \bibfield  {author} {\bibinfo {author} {\bibfnamefont {L.}~\bibnamefont
  {\ifmmode~\check{S}\else \v{S}\fi{}mejkal}}, \bibinfo {author} {\bibfnamefont
  {A.~B.}\ \bibnamefont {Hellenes}}, \bibinfo {author} {\bibfnamefont
  {R.}~\bibnamefont {Gonz\'alez-Hern\'andez}}, \bibinfo {author} {\bibfnamefont
  {J.}~\bibnamefont {Sinova}},\ and\ \bibinfo {author} {\bibfnamefont
  {T.}~\bibnamefont {Jungwirth}},\ }\bibfield  {title} {\bibinfo {title} {Giant
  and tunneling magnetoresistance in unconventional collinear antiferromagnets
  with nonrelativistic spin-momentum coupling},\ }\href
  {https://doi.org/10.1103/PhysRevX.12.011028} {\bibfield  {journal} {\bibinfo
  {journal} {Phys. Rev. X}\ }\textbf {\bibinfo {volume} {12}},\ \bibinfo
  {pages} {011028} (\bibinfo {year} {2022}{\natexlab{c}})}\BibitemShut
  {NoStop}%
\bibitem [{\citenamefont {Shao}\ \emph {et~al.}(2021)\citenamefont {Shao},
  \citenamefont {Zhang}, \citenamefont {Li}, \citenamefont {Eom},\ and\
  \citenamefont {Tsymbal}}]{Shao2021}%
  \BibitemOpen
  \bibfield  {author} {\bibinfo {author} {\bibfnamefont {D.-F.}\ \bibnamefont
  {Shao}}, \bibinfo {author} {\bibfnamefont {S.-H.}\ \bibnamefont {Zhang}},
  \bibinfo {author} {\bibfnamefont {M.}~\bibnamefont {Li}}, \bibinfo {author}
  {\bibfnamefont {C.-B.}\ \bibnamefont {Eom}},\ and\ \bibinfo {author}
  {\bibfnamefont {E.~Y.}\ \bibnamefont {Tsymbal}},\ }\bibfield  {title}
  {\bibinfo {title} {Spin-neutral currents for spintronics},\ }\href
  {https://doi.org/10.1038/s41467-021-26915-3} {\bibfield  {journal} {\bibinfo
  {journal} {Nat. Commun.}\ }\textbf {\bibinfo {volume} {12}},\ \bibinfo
  {pages} {7061} (\bibinfo {year} {2021})}\BibitemShut {NoStop}%
\bibitem [{\citenamefont {Chen}\ \emph {et~al.}(2024)\citenamefont {Chen},
  \citenamefont {Wang}, \citenamefont {Qin}, \citenamefont {Meng},
  \citenamefont {Zhou}, \citenamefont {Wang}, \citenamefont {Liu},
  \citenamefont {Zhao}, \citenamefont {Duan}, \citenamefont {Zhang},
  \citenamefont {Liu}, \citenamefont {Shao},\ and\ \citenamefont
  {Liu}}]{Chen2024}%
  \BibitemOpen
  \bibfield  {author} {\bibinfo {author} {\bibfnamefont {H.}~\bibnamefont
  {Chen}}, \bibinfo {author} {\bibfnamefont {Z.}~\bibnamefont {Wang}}, \bibinfo
  {author} {\bibfnamefont {P.}~\bibnamefont {Qin}}, \bibinfo {author}
  {\bibfnamefont {Z.}~\bibnamefont {Meng}}, \bibinfo {author} {\bibfnamefont
  {X.}~\bibnamefont {Zhou}}, \bibinfo {author} {\bibfnamefont {X.}~\bibnamefont
  {Wang}}, \bibinfo {author} {\bibfnamefont {L.}~\bibnamefont {Liu}}, \bibinfo
  {author} {\bibfnamefont {G.}~\bibnamefont {Zhao}}, \bibinfo {author}
  {\bibfnamefont {Z.}~\bibnamefont {Duan}}, \bibinfo {author} {\bibfnamefont
  {T.}~\bibnamefont {Zhang}}, \bibinfo {author} {\bibfnamefont
  {J.}~\bibnamefont {Liu}}, \bibinfo {author} {\bibfnamefont {D.}~\bibnamefont
  {Shao}},\ and\ \bibinfo {author} {\bibfnamefont {Z.}~\bibnamefont {Liu}},\
  }\bibfield  {title} {\bibinfo {title} {Altermagnetic spin-splitting
  magnetoresistance},\ }\href {https://arxiv.org/pdf/2412.18220.pdf} {\bibfield
   {journal} {\bibinfo  {journal} {arXiv}\ } (\bibinfo {year} {2024})},\
  \Eprint {https://arxiv.org/abs/2412.18220} {2412.18220} \BibitemShut
  {NoStop}%
\bibitem [{\citenamefont {Gonzalez~Betancourt}\ \emph
  {et~al.}(2024)\citenamefont {Gonzalez~Betancourt}, \citenamefont {Zub{\'a}{\v
  c}}, \citenamefont {Geishendorf}, \citenamefont {Ritzinger}, \citenamefont
  {R{\r u}{\v z}i{\v c}kov{\'a}}, \citenamefont {Kotte}, \citenamefont {{\v
  Z}elezn{\'y}}, \citenamefont {Olejn{\'\i}k}, \citenamefont {Springholz},
  \citenamefont {B{\"u}chner}, \citenamefont {Thomas}, \citenamefont
  {V{\'y}born{\'y}}, \citenamefont {Jungwirth}, \citenamefont {Reichlov{\'a}},\
  and\ \citenamefont {Kriegner}}]{Betancourt2024}%
  \BibitemOpen
  \bibfield  {author} {\bibinfo {author} {\bibfnamefont {R.~D.}\ \bibnamefont
  {Gonzalez~Betancourt}}, \bibinfo {author} {\bibfnamefont {J.}~\bibnamefont
  {Zub{\'a}{\v c}}}, \bibinfo {author} {\bibfnamefont {K.}~\bibnamefont
  {Geishendorf}}, \bibinfo {author} {\bibfnamefont {P.}~\bibnamefont
  {Ritzinger}}, \bibinfo {author} {\bibfnamefont {B.}~\bibnamefont {R{\r u}{\v
  z}i{\v c}kov{\'a}}}, \bibinfo {author} {\bibfnamefont {T.}~\bibnamefont
  {Kotte}}, \bibinfo {author} {\bibfnamefont {J.}~\bibnamefont {{\v
  Z}elezn{\'y}}}, \bibinfo {author} {\bibfnamefont {K.}~\bibnamefont
  {Olejn{\'\i}k}}, \bibinfo {author} {\bibfnamefont {G.}~\bibnamefont
  {Springholz}}, \bibinfo {author} {\bibfnamefont {B.}~\bibnamefont
  {B{\"u}chner}}, \bibinfo {author} {\bibfnamefont {A.}~\bibnamefont {Thomas}},
  \bibinfo {author} {\bibfnamefont {K.}~\bibnamefont {V{\'y}born{\'y}}},
  \bibinfo {author} {\bibfnamefont {T.}~\bibnamefont {Jungwirth}}, \bibinfo
  {author} {\bibfnamefont {H.}~\bibnamefont {Reichlov{\'a}}},\ and\ \bibinfo
  {author} {\bibfnamefont {D.}~\bibnamefont {Kriegner}},\ }\bibfield  {title}
  {\bibinfo {title} {Anisotropic magnetoresistance in altermagnetic
  \text{MnTe}},\ }\href {https://doi.org/10.1038/s44306-024-00046-z} {\bibfield
   {journal} {\bibinfo  {journal} {npj Spintronics}\ }\textbf {\bibinfo
  {volume} {2}},\ \bibinfo {pages} {45} (\bibinfo {year} {2024})}\BibitemShut
  {NoStop}%
\bibitem [{\citenamefont {Peng}\ \emph {et~al.}(2024)\citenamefont {Peng},
  \citenamefont {Wang}, \citenamefont {Zhang}, \citenamefont {Zhou},
  \citenamefont {Sun}, \citenamefont {Su}, \citenamefont {Wu}, \citenamefont
  {Zhou}, \citenamefont {Liu}, \citenamefont {Wang}, \citenamefont {Yang},
  \citenamefont {Chen}, \citenamefont {Fang}, \citenamefont {Du}, \citenamefont
  {Jiao}, \citenamefont {Wu},\ and\ \citenamefont {Fang}}]{Peng2024}%
  \BibitemOpen
  \bibfield  {author} {\bibinfo {author} {\bibfnamefont {X.}~\bibnamefont
  {Peng}}, \bibinfo {author} {\bibfnamefont {Y.}~\bibnamefont {Wang}}, \bibinfo
  {author} {\bibfnamefont {S.}~\bibnamefont {Zhang}}, \bibinfo {author}
  {\bibfnamefont {Y.}~\bibnamefont {Zhou}}, \bibinfo {author} {\bibfnamefont
  {Y.}~\bibnamefont {Sun}}, \bibinfo {author} {\bibfnamefont {Y.}~\bibnamefont
  {Su}}, \bibinfo {author} {\bibfnamefont {C.}~\bibnamefont {Wu}}, \bibinfo
  {author} {\bibfnamefont {T.}~\bibnamefont {Zhou}}, \bibinfo {author}
  {\bibfnamefont {L.}~\bibnamefont {Liu}}, \bibinfo {author} {\bibfnamefont
  {H.}~\bibnamefont {Wang}}, \bibinfo {author} {\bibfnamefont {J.}~\bibnamefont
  {Yang}}, \bibinfo {author} {\bibfnamefont {B.}~\bibnamefont {Chen}}, \bibinfo
  {author} {\bibfnamefont {Z.}~\bibnamefont {Fang}}, \bibinfo {author}
  {\bibfnamefont {J.}~\bibnamefont {Du}}, \bibinfo {author} {\bibfnamefont
  {Z.}~\bibnamefont {Jiao}}, \bibinfo {author} {\bibfnamefont {Q.}~\bibnamefont
  {Wu}},\ and\ \bibinfo {author} {\bibfnamefont {M.}~\bibnamefont {Fang}},\
  }\bibfield  {title} {\bibinfo {title} {Scaling behavior of magnetoresistance
  and \text{Hall} resistivity in altermagnet \text{CrSb}},\ }\href
  {https://arxiv.org/pdf/2412.12263v1.pdf} {\bibfield  {journal} {\bibinfo
  {journal} {arXiv}\ } (\bibinfo {year} {2024})},\ \Eprint
  {https://arxiv.org/abs/2412.12263} {2412.12263} \BibitemShut {NoStop}%
\bibitem [{\citenamefont {Liu}\ \emph {et~al.}(2024{\natexlab{a}})\citenamefont
  {Liu}, \citenamefont {Zhang}, \citenamefont {Yuan}, \citenamefont {Liu},
  \citenamefont {Zhu}, \citenamefont {Lu},\ and\ \citenamefont
  {Xiong}}]{Liu2024}%
  \BibitemOpen
  \bibfield  {author} {\bibinfo {author} {\bibfnamefont {F.}~\bibnamefont
  {Liu}}, \bibinfo {author} {\bibfnamefont {Z.}~\bibnamefont {Zhang}}, \bibinfo
  {author} {\bibfnamefont {X.}~\bibnamefont {Yuan}}, \bibinfo {author}
  {\bibfnamefont {Y.}~\bibnamefont {Liu}}, \bibinfo {author} {\bibfnamefont
  {S.}~\bibnamefont {Zhu}}, \bibinfo {author} {\bibfnamefont {Z.}~\bibnamefont
  {Lu}},\ and\ \bibinfo {author} {\bibfnamefont {R.}~\bibnamefont {Xiong}},\
  }\bibfield  {title} {\bibinfo {title} {Giant tunneling magnetoresistance in
  insulated altermagnet/ferromagnet junctions induced by spin-dependent
  tunneling effect},\ }\href {https://doi.org/10.1103/PhysRevB.110.134437}
  {\bibfield  {journal} {\bibinfo  {journal} {Phys. Rev. B}\ }\textbf {\bibinfo
  {volume} {110}},\ \bibinfo {pages} {134437} (\bibinfo {year}
  {2024}{\natexlab{a}})}\BibitemShut {NoStop}%
\bibitem [{\citenamefont {Gonzalez~Betancourt}\ \emph
  {et~al.}(2023)\citenamefont {Gonzalez~Betancourt}, \citenamefont
  {Zub\'a\ifmmode~\check{c}\else \v{c}\fi{}}, \citenamefont
  {Gonzalez-Hernandez}, \citenamefont {Geishendorf}, \citenamefont {\ifmmode
  \check{S}\else \v{S}\fi{}ob\'a\ifmmode~\check{n}\else \v{n}\fi{}},
  \citenamefont {Springholz}, \citenamefont {Olejn\'{\i}k}, \citenamefont
  {\ifmmode~\check{S}\else \v{S}\fi{}mejkal}, \citenamefont {Sinova},
  \citenamefont {Jungwirth}, \citenamefont {Goennenwein}, \citenamefont
  {Thomas}, \citenamefont {Reichlov\'a}, \citenamefont {\ifmmode~\check{Z}\else
  \v{Z}\fi{}elezn\'y},\ and\ \citenamefont {Kriegner}}]{Betancourt2023}%
  \BibitemOpen
  \bibfield  {author} {\bibinfo {author} {\bibfnamefont {R.~D.}\ \bibnamefont
  {Gonzalez~Betancourt}}, \bibinfo {author} {\bibfnamefont {J.}~\bibnamefont
  {Zub\'a\ifmmode~\check{c}\else \v{c}\fi{}}}, \bibinfo {author} {\bibfnamefont
  {R.}~\bibnamefont {Gonzalez-Hernandez}}, \bibinfo {author} {\bibfnamefont
  {K.}~\bibnamefont {Geishendorf}}, \bibinfo {author} {\bibfnamefont
  {Z.}~\bibnamefont {\ifmmode \check{S}\else
  \v{S}\fi{}ob\'a\ifmmode~\check{n}\else \v{n}\fi{}}}, \bibinfo {author}
  {\bibfnamefont {G.}~\bibnamefont {Springholz}}, \bibinfo {author}
  {\bibfnamefont {K.}~\bibnamefont {Olejn\'{\i}k}}, \bibinfo {author}
  {\bibfnamefont {L.}~\bibnamefont {\ifmmode~\check{S}\else \v{S}\fi{}mejkal}},
  \bibinfo {author} {\bibfnamefont {J.}~\bibnamefont {Sinova}}, \bibinfo
  {author} {\bibfnamefont {T.}~\bibnamefont {Jungwirth}}, \bibinfo {author}
  {\bibfnamefont {S.~T.~B.}\ \bibnamefont {Goennenwein}}, \bibinfo {author}
  {\bibfnamefont {A.}~\bibnamefont {Thomas}}, \bibinfo {author} {\bibfnamefont
  {H.}~\bibnamefont {Reichlov\'a}}, \bibinfo {author} {\bibfnamefont
  {J.}~\bibnamefont {\ifmmode~\check{Z}\else \v{Z}\fi{}elezn\'y}},\ and\
  \bibinfo {author} {\bibfnamefont {D.}~\bibnamefont {Kriegner}},\ }\bibfield
  {title} {\bibinfo {title} {Spontaneous anomalous \text{H}all effect arising
  from an unconventional compensated magnetic phase in a semiconductor},\
  }\href {https://doi.org/10.1103/PhysRevLett.130.036702} {\bibfield  {journal}
  {\bibinfo  {journal} {Phys. Rev. Lett.}\ }\textbf {\bibinfo {volume} {130}},\
  \bibinfo {pages} {036702} (\bibinfo {year} {2023})}\BibitemShut {NoStop}%
\bibitem [{\citenamefont {Reichlova}\ \emph {et~al.}(2024)\citenamefont
  {Reichlova}, \citenamefont {Lopes~Seeger}, \citenamefont
  {Gonz{\'a}lez-Hern{\'a}ndez}, \citenamefont {Kounta}, \citenamefont
  {Schlitz}, \citenamefont {Kriegner}, \citenamefont {Ritzinger}, \citenamefont
  {Lammel}, \citenamefont {Leivisk{\"a}}, \citenamefont {Birk~Hellenes},
  \citenamefont {Olejn{\'\i}k}, \citenamefont {Pet{\v r}i{\v c}ek},
  \citenamefont {Dole{\v z}al}, \citenamefont {Horak}, \citenamefont
  {Schmoranzerova}, \citenamefont {Badura}, \citenamefont {Bertaina},
  \citenamefont {Thomas}, \citenamefont {Baltz}, \citenamefont {Michez},
  \citenamefont {Sinova}, \citenamefont {Goennenwein}, \citenamefont
  {Jungwirth},\ and\ \citenamefont {{\v S}mejkal}}]{Reichlova2024}%
  \BibitemOpen
  \bibfield  {author} {\bibinfo {author} {\bibfnamefont {H.}~\bibnamefont
  {Reichlova}}, \bibinfo {author} {\bibfnamefont {R.}~\bibnamefont
  {Lopes~Seeger}}, \bibinfo {author} {\bibfnamefont {R.}~\bibnamefont
  {Gonz{\'a}lez-Hern{\'a}ndez}}, \bibinfo {author} {\bibfnamefont
  {I.}~\bibnamefont {Kounta}}, \bibinfo {author} {\bibfnamefont
  {R.}~\bibnamefont {Schlitz}}, \bibinfo {author} {\bibfnamefont
  {D.}~\bibnamefont {Kriegner}}, \bibinfo {author} {\bibfnamefont
  {P.}~\bibnamefont {Ritzinger}}, \bibinfo {author} {\bibfnamefont
  {M.}~\bibnamefont {Lammel}}, \bibinfo {author} {\bibfnamefont
  {M.}~\bibnamefont {Leivisk{\"a}}}, \bibinfo {author} {\bibfnamefont
  {A.}~\bibnamefont {Birk~Hellenes}}, \bibinfo {author} {\bibfnamefont
  {K.}~\bibnamefont {Olejn{\'\i}k}}, \bibinfo {author} {\bibfnamefont
  {V.}~\bibnamefont {Pet{\v r}i{\v c}ek}}, \bibinfo {author} {\bibfnamefont
  {P.}~\bibnamefont {Dole{\v z}al}}, \bibinfo {author} {\bibfnamefont
  {L.}~\bibnamefont {Horak}}, \bibinfo {author} {\bibfnamefont
  {E.}~\bibnamefont {Schmoranzerova}}, \bibinfo {author} {\bibfnamefont
  {A.}~\bibnamefont {Badura}}, \bibinfo {author} {\bibfnamefont
  {S.}~\bibnamefont {Bertaina}}, \bibinfo {author} {\bibfnamefont
  {A.}~\bibnamefont {Thomas}}, \bibinfo {author} {\bibfnamefont
  {V.}~\bibnamefont {Baltz}}, \bibinfo {author} {\bibfnamefont
  {L.}~\bibnamefont {Michez}}, \bibinfo {author} {\bibfnamefont
  {J.}~\bibnamefont {Sinova}}, \bibinfo {author} {\bibfnamefont {S.~T.~B.}\
  \bibnamefont {Goennenwein}}, \bibinfo {author} {\bibfnamefont
  {T.}~\bibnamefont {Jungwirth}},\ and\ \bibinfo {author} {\bibfnamefont
  {L.}~\bibnamefont {{\v S}mejkal}},\ }\bibfield  {title} {\bibinfo {title}
  {Observation of a spontaneous anomalous \text{H}all response in the
  \text{Mn}$_5$\text{Si}$_3$ d-wave altermagnet candidate},\ }\href
  {https://doi.org/10.1038/s41467-024-48493-w} {\bibfield  {journal} {\bibinfo
  {journal} {Nat. Commun.}\ }\textbf {\bibinfo {volume} {15}},\ \bibinfo
  {pages} {4961} (\bibinfo {year} {2024})}\BibitemShut {NoStop}%
\bibitem [{\citenamefont {Wadley}\ \emph {et~al.}(2016)\citenamefont {Wadley},
  \citenamefont {Howells}, \citenamefont {{\v{Z}}elezn{\`y}}, \citenamefont
  {Andrews}, \citenamefont {Hills}, \citenamefont {Campion}, \citenamefont
  {Nov{\'a}k}, \citenamefont {Olejn{\'\i}k}, \citenamefont {Maccherozzi},
  \citenamefont {Dhesi} \emph {et~al.}}]{wadley2016electrical}%
  \BibitemOpen
  \bibfield  {author} {\bibinfo {author} {\bibfnamefont {P.}~\bibnamefont
  {Wadley}}, \bibinfo {author} {\bibfnamefont {B.}~\bibnamefont {Howells}},
  \bibinfo {author} {\bibfnamefont {J.}~\bibnamefont {{\v{Z}}elezn{\`y}}},
  \bibinfo {author} {\bibfnamefont {C.}~\bibnamefont {Andrews}}, \bibinfo
  {author} {\bibfnamefont {V.}~\bibnamefont {Hills}}, \bibinfo {author}
  {\bibfnamefont {R.~P.}\ \bibnamefont {Campion}}, \bibinfo {author}
  {\bibfnamefont {V.}~\bibnamefont {Nov{\'a}k}}, \bibinfo {author}
  {\bibfnamefont {K.}~\bibnamefont {Olejn{\'\i}k}}, \bibinfo {author}
  {\bibfnamefont {F.}~\bibnamefont {Maccherozzi}}, \bibinfo {author}
  {\bibfnamefont {S.}~\bibnamefont {Dhesi}}, \emph {et~al.},\ }\bibfield
  {title} {\bibinfo {title} {Electrical switching of an antiferromagnet},\
  }\href {https://doi.org/DOI: 10.1126/science.aab1031} {\bibfield  {journal}
  {\bibinfo  {journal} {Science}\ }\textbf {\bibinfo {volume} {351}},\ \bibinfo
  {pages} {587} (\bibinfo {year} {2016})}\BibitemShut {NoStop}%
\bibitem [{\citenamefont {Godinho}\ \emph {et~al.}(2018)\citenamefont
  {Godinho}, \citenamefont {Reichlov{\'a}}, \citenamefont {Kriegner},
  \citenamefont {Nov{\'a}k}, \citenamefont {Olejn{\'{\i}}k}, \citenamefont
  {Ka{\v s}par}, \citenamefont {{\v S}ob{\'a}{\v n}}, \citenamefont {Wadley},
  \citenamefont {Campion}, \citenamefont {Otxoa}, \citenamefont {Roy},
  \citenamefont {{\v Z}elezn{\'y}}, \citenamefont {Jungwirth},\ and\
  \citenamefont {Wunderlich}}]{Godinho2018}%
  \BibitemOpen
  \bibfield  {author} {\bibinfo {author} {\bibfnamefont {J.}~\bibnamefont
  {Godinho}}, \bibinfo {author} {\bibfnamefont {H.}~\bibnamefont
  {Reichlov{\'a}}}, \bibinfo {author} {\bibfnamefont {D.}~\bibnamefont
  {Kriegner}}, \bibinfo {author} {\bibfnamefont {V.}~\bibnamefont {Nov{\'a}k}},
  \bibinfo {author} {\bibfnamefont {K.}~\bibnamefont {Olejn{\'{\i}}k}},
  \bibinfo {author} {\bibfnamefont {Z.}~\bibnamefont {Ka{\v s}par}}, \bibinfo
  {author} {\bibfnamefont {Z.}~\bibnamefont {{\v S}ob{\'a}{\v n}}}, \bibinfo
  {author} {\bibfnamefont {P.}~\bibnamefont {Wadley}}, \bibinfo {author}
  {\bibfnamefont {R.~P.}\ \bibnamefont {Campion}}, \bibinfo {author}
  {\bibfnamefont {R.~M.}\ \bibnamefont {Otxoa}}, \bibinfo {author}
  {\bibfnamefont {P.~E.}\ \bibnamefont {Roy}}, \bibinfo {author} {\bibfnamefont
  {J.}~\bibnamefont {{\v Z}elezn{\'y}}}, \bibinfo {author} {\bibfnamefont
  {T.}~\bibnamefont {Jungwirth}},\ and\ \bibinfo {author} {\bibfnamefont
  {J.}~\bibnamefont {Wunderlich}},\ }\bibfield  {title} {\bibinfo {title}
  {Electrically induced and detected n{\'e}el vector reversal in a collinear
  antiferromagnet},\ }\href {https://doi.org/10.1038/s41467-018-07092-2}
  {\bibfield  {journal} {\bibinfo  {journal} {Nat. Commun.}\ }\textbf {\bibinfo
  {volume} {9}},\ \bibinfo {eid} {4686} (\bibinfo {year} {2018})}\BibitemShut
  {NoStop}%
\bibitem [{\citenamefont {Sun}\ \emph {et~al.}(2023)\citenamefont {Sun},
  \citenamefont {Brataas},\ and\ \citenamefont {Linder}}]{sun2023dndreev}%
  \BibitemOpen
  \bibfield  {author} {\bibinfo {author} {\bibfnamefont {C.}~\bibnamefont
  {Sun}}, \bibinfo {author} {\bibfnamefont {A.}~\bibnamefont {Brataas}},\ and\
  \bibinfo {author} {\bibfnamefont {J.}~\bibnamefont {Linder}},\ }\bibfield
  {title} {\bibinfo {title} {Andreev reflection in altermagnets},\ }\href
  {https://doi.org/10.1103/PhysRevB.108.054511} {\bibfield  {journal} {\bibinfo
   {journal} {Phys. Rev. B}\ }\textbf {\bibinfo {volume} {108}},\ \bibinfo
  {pages} {054511} (\bibinfo {year} {2023})}\BibitemShut {NoStop}%
\bibitem [{\citenamefont {Beenakker}\ and\ \citenamefont
  {Vakhtel}(2023)}]{beenakker2024phase}%
  \BibitemOpen
  \bibfield  {author} {\bibinfo {author} {\bibfnamefont {C.~W.~J.}\
  \bibnamefont {Beenakker}}\ and\ \bibinfo {author} {\bibfnamefont
  {T.}~\bibnamefont {Vakhtel}},\ }\bibfield  {title} {\bibinfo {title}
  {Phase-shifted andreev levels in an altermagnet josephson junction},\ }\href
  {https://doi.org/10.1103/PhysRevB.108.075425} {\bibfield  {journal} {\bibinfo
   {journal} {Phys. Rev. B}\ }\textbf {\bibinfo {volume} {108}},\ \bibinfo
  {pages} {075425} (\bibinfo {year} {2023})}\BibitemShut {NoStop}%
\bibitem [{\citenamefont {Winkler}\ \emph {et~al.}(2003)\citenamefont
  {Winkler}, \citenamefont {Papadakis}, \citenamefont {De~Poortere},\ and\
  \citenamefont {Shayegan}}]{winkler2003spin}%
  \BibitemOpen
  \bibfield  {author} {\bibinfo {author} {\bibfnamefont {R.}~\bibnamefont
  {Winkler}}, \bibinfo {author} {\bibfnamefont {S.}~\bibnamefont {Papadakis}},
  \bibinfo {author} {\bibfnamefont {E.}~\bibnamefont {De~Poortere}},\ and\
  \bibinfo {author} {\bibfnamefont {M.}~\bibnamefont {Shayegan}},\ }\href@noop
  {} {\emph {\bibinfo {title} {Spin-Orbit Coupling in Two-Dimensional Electron
  and Hole Systems}}},\ Vol.~\bibinfo {volume} {41}\ (\bibinfo  {publisher}
  {Springer},\ \bibinfo {year} {2003})\BibitemShut {NoStop}%
\bibitem [{\citenamefont {Messiah}(2014)}]{messiah2014quantum}%
  \BibitemOpen
  \bibfield  {author} {\bibinfo {author} {\bibfnamefont {A.}~\bibnamefont
  {Messiah}},\ }\href@noop {} {\emph {\bibinfo {title} {Quantum mechanics}}}\
  (\bibinfo  {publisher} {Courier Corporation},\ \bibinfo {year}
  {2014})\BibitemShut {NoStop}%
\bibitem [{\citenamefont {Zhang}\ \emph {et~al.}(2019)\citenamefont {Zhang},
  \citenamefont {Bergeret},\ and\ \citenamefont {Golovach}}]{zhang2019theory}%
  \BibitemOpen
  \bibfield  {author} {\bibinfo {author} {\bibfnamefont {X.-P.}\ \bibnamefont
  {Zhang}}, \bibinfo {author} {\bibfnamefont {F.~S.}\ \bibnamefont
  {Bergeret}},\ and\ \bibinfo {author} {\bibfnamefont {V.~N.}\ \bibnamefont
  {Golovach}},\ }\bibfield  {title} {\bibinfo {title} {Theory of spin
  \text{H}all magnetoresistance from a microscopic perspective},\ }\href
  {https://doi.org/https://doi.org/10.1021/acs.nanolett.9b02459} {\bibfield
  {journal} {\bibinfo  {journal} {Nano Lett.}\ }\textbf {\bibinfo {volume}
  {19}},\ \bibinfo {pages} {6330} (\bibinfo {year} {2019})}\BibitemShut
  {NoStop}%
\bibitem [{\citenamefont {Zhang}(2022)}]{zhang2023extrinsic}%
  \BibitemOpen
  \bibfield  {author} {\bibinfo {author} {\bibfnamefont {X.-P.}\ \bibnamefont
  {Zhang}},\ }\bibfield  {title} {\bibinfo {title} {Extrinsic spin-valley
  \text{H}all effect and spin-relaxation anisotropy in magnetized and strained
  graphene},\ }\href {https://doi.org/10.1103/PhysRevB.106.115437} {\bibfield
  {journal} {\bibinfo  {journal} {Phys. Rev. B}\ }\textbf {\bibinfo {volume}
  {106}},\ \bibinfo {pages} {115437} (\bibinfo {year} {2022})}\BibitemShut
  {NoStop}%
\bibitem [{\citenamefont {Zhang}\ \emph {et~al.}(2023)\citenamefont {Zhang},
  \citenamefont {Yao}, \citenamefont {Wang},\ and\ \citenamefont
  {Yan}}]{zhang2022microscopic}%
  \BibitemOpen
  \bibfield  {author} {\bibinfo {author} {\bibfnamefont {X.-P.}\ \bibnamefont
  {Zhang}}, \bibinfo {author} {\bibfnamefont {Y.}~\bibnamefont {Yao}}, \bibinfo
  {author} {\bibfnamefont {K.~Y.}\ \bibnamefont {Wang}},\ and\ \bibinfo
  {author} {\bibfnamefont {P.}~\bibnamefont {Yan}},\ }\bibfield  {title}
  {\bibinfo {title} {Microscopic theory of spin-orbit torque and spin memory
  loss from interfacial spin-orbit coupling},\ }\href
  {https://doi.org/10.1103/PhysRevB.108.125309} {\bibfield  {journal} {\bibinfo
   {journal} {Phys. Rev. B}\ }\textbf {\bibinfo {volume} {108}},\ \bibinfo
  {pages} {125309} (\bibinfo {year} {2023})}\BibitemShut {NoStop}%
\bibitem [{Note1()}]{Note1}%
  \BibitemOpen
  \bibinfo {note} {The SEC for Co adatoms on the Cu(100) surface has been
  theoretically predicted to reach a substantial ferromagnetic interaction of
  approximately $\protect \mathcal {J}n_s \simeq 350$ meV~\cite
  {wahl2007exchange}. Although experimental data for SEC in the altermagnets
  CrSb and RuO$_2$ are currently unavailable, we consider $\protect \mathcal
  {J}n_s \sim $ meV to be a physically reasonable estimate.}\BibitemShut
  {Stop}%
\bibitem [{\citenamefont {Zhang}\ \emph {et~al.}(2024)\citenamefont {Zhang},
  \citenamefont {Wang},\ and\ \citenamefont {Yao}}]{zhang2024microscopic}%
  \BibitemOpen
  \bibfield  {author} {\bibinfo {author} {\bibfnamefont {X.-P.}\ \bibnamefont
  {Zhang}}, \bibinfo {author} {\bibfnamefont {X.}~\bibnamefont {Wang}},\ and\
  \bibinfo {author} {\bibfnamefont {Y.}~\bibnamefont {Yao}},\ }\bibfield
  {title} {\bibinfo {title} {Microscopic theory of magnetoresistance in
  ferromagnetic materials},\ }\href {https://doi.org/10.48550/arXiv.2406.13932}
  {\bibfield  {journal} {\bibinfo  {journal} {arXiv}\ } (\bibinfo {year}
  {2024})},\ \Eprint {https://arxiv.org/abs/2406.13932} {2406.13932}
  \BibitemShut {NoStop}%
\bibitem [{\citenamefont {Breuer}\ and\ \citenamefont
  {Petruccione}(2002)}]{breuer2002theory}%
  \BibitemOpen
  \bibfield  {author} {\bibinfo {author} {\bibfnamefont {H.-P.}\ \bibnamefont
  {Breuer}}\ and\ \bibinfo {author} {\bibfnamefont {F.}~\bibnamefont
  {Petruccione}},\ }\href@noop {} {\emph {\bibinfo {title} {The theory of open
  quantum systems}}}\ (\bibinfo  {publisher} {OUP Oxford},\ \bibinfo {year}
  {2002})\BibitemShut {NoStop}%
\bibitem [{\citenamefont {Liao}\ \emph {et~al.}(2024)\citenamefont {Liao},
  \citenamefont {Wang}, \citenamefont {Tien}, \citenamefont {Huang},\ and\
  \citenamefont {Qu}}]{liao2024separation}%
  \BibitemOpen
  \bibfield  {author} {\bibinfo {author} {\bibfnamefont {C.-T.}\ \bibnamefont
  {Liao}}, \bibinfo {author} {\bibfnamefont {Y.-C.}\ \bibnamefont {Wang}},
  \bibinfo {author} {\bibfnamefont {Y.-C.}\ \bibnamefont {Tien}}, \bibinfo
  {author} {\bibfnamefont {S.-Y.}\ \bibnamefont {Huang}},\ and\ \bibinfo
  {author} {\bibfnamefont {D.}~\bibnamefont {Qu}},\ }\bibfield  {title}
  {\bibinfo {title} {Separation of inverse altermagnetic spin-splitting effect
  from inverse spin \text{H}all effect in \text{RuO}$_{2}$},\ }\href
  {https://doi.org/10.1103/PhysRevLett.133.056701} {\bibfield  {journal}
  {\bibinfo  {journal} {Phys. Rev. Lett.}\ }\textbf {\bibinfo {volume} {133}},\
  \bibinfo {pages} {056701} (\bibinfo {year} {2024})}\BibitemShut {NoStop}%
\bibitem [{\citenamefont {Feng}\ \emph {et~al.}(2024)\citenamefont {Feng},
  \citenamefont {Bai}, \citenamefont {Fan}, \citenamefont {Guo}, \citenamefont
  {Zhang}, \citenamefont {Chai}, \citenamefont {Wang}, \citenamefont {Xue},
  \citenamefont {Song},\ and\ \citenamefont {Fan}}]{feng2024incommensurate}%
  \BibitemOpen
  \bibfield  {author} {\bibinfo {author} {\bibfnamefont {X.}~\bibnamefont
  {Feng}}, \bibinfo {author} {\bibfnamefont {H.}~\bibnamefont {Bai}}, \bibinfo
  {author} {\bibfnamefont {X.}~\bibnamefont {Fan}}, \bibinfo {author}
  {\bibfnamefont {M.}~\bibnamefont {Guo}}, \bibinfo {author} {\bibfnamefont
  {Z.}~\bibnamefont {Zhang}}, \bibinfo {author} {\bibfnamefont
  {G.}~\bibnamefont {Chai}}, \bibinfo {author} {\bibfnamefont {T.}~\bibnamefont
  {Wang}}, \bibinfo {author} {\bibfnamefont {D.}~\bibnamefont {Xue}}, \bibinfo
  {author} {\bibfnamefont {C.}~\bibnamefont {Song}},\ and\ \bibinfo {author}
  {\bibfnamefont {X.}~\bibnamefont {Fan}},\ }\bibfield  {title} {\bibinfo
  {title} {Incommensurate spin density wave in antiferromagnetic
  \text{RuO}$_{2}$ evinced by abnormal spin splitting torque},\ }\href
  {https://doi.org/10.1103/PhysRevLett.132.086701} {\bibfield  {journal}
  {\bibinfo  {journal} {Phys. Rev. Lett.}\ }\textbf {\bibinfo {volume} {132}},\
  \bibinfo {pages} {086701} (\bibinfo {year} {2024})}\BibitemShut {NoStop}%
\bibitem [{\citenamefont {Chen}\ \emph {et~al.}(2013)\citenamefont {Chen},
  \citenamefont {Takahashi}, \citenamefont {Nakayama}, \citenamefont
  {Althammer}, \citenamefont {Goennenwein}, \citenamefont {Saitoh},\ and\
  \citenamefont {Bauer}}]{chen2013theory}%
  \BibitemOpen
  \bibfield  {author} {\bibinfo {author} {\bibfnamefont {Y.-T.}\ \bibnamefont
  {Chen}}, \bibinfo {author} {\bibfnamefont {S.}~\bibnamefont {Takahashi}},
  \bibinfo {author} {\bibfnamefont {H.}~\bibnamefont {Nakayama}}, \bibinfo
  {author} {\bibfnamefont {M.}~\bibnamefont {Althammer}}, \bibinfo {author}
  {\bibfnamefont {S.~T.~B.}\ \bibnamefont {Goennenwein}}, \bibinfo {author}
  {\bibfnamefont {E.}~\bibnamefont {Saitoh}},\ and\ \bibinfo {author}
  {\bibfnamefont {G.~E.~W.}\ \bibnamefont {Bauer}},\ }\bibfield  {title}
  {\bibinfo {title} {Theory of spin \text{H}all magnetoresistance},\ }\href
  {https://doi.org/10.1103/PhysRevB.87.144411} {\bibfield  {journal} {\bibinfo
  {journal} {Phys. Rev. B}\ }\textbf {\bibinfo {volume} {87}},\ \bibinfo
  {pages} {144411} (\bibinfo {year} {2013})}\BibitemShut {NoStop}%
\bibitem [{\citenamefont {Maekawa}\ and\ \citenamefont
  {Kimura}(2017)}]{maekawa2017spin}%
  \BibitemOpen
  \bibfield  {author} {\bibinfo {author} {\bibfnamefont {S.}~\bibnamefont
  {Maekawa}}\ and\ \bibinfo {author} {\bibfnamefont {T.}~\bibnamefont
  {Kimura}},\ }\href@noop {} {\emph {\bibinfo {title} {Spin Current}}},\
  Vol.~\bibinfo {volume} {22}\ (\bibinfo  {publisher} {Oxford University
  Press},\ \bibinfo {year} {2017})\BibitemShut {NoStop}%
\bibitem [{\citenamefont {Sinova}\ \emph {et~al.}(2015)\citenamefont {Sinova},
  \citenamefont {Valenzuela}, \citenamefont {Wunderlich}, \citenamefont
  {Back},\ and\ \citenamefont {Jungwirth}}]{sinova2015spin}%
  \BibitemOpen
  \bibfield  {author} {\bibinfo {author} {\bibfnamefont {J.}~\bibnamefont
  {Sinova}}, \bibinfo {author} {\bibfnamefont {S.~O.}\ \bibnamefont
  {Valenzuela}}, \bibinfo {author} {\bibfnamefont {J.}~\bibnamefont
  {Wunderlich}}, \bibinfo {author} {\bibfnamefont {C.~H.}\ \bibnamefont
  {Back}},\ and\ \bibinfo {author} {\bibfnamefont {T.}~\bibnamefont
  {Jungwirth}},\ }\bibfield  {title} {\bibinfo {title} {Spin \text{H}all
  effects},\ }\href {https://doi.org/10.1103/RevModPhys.87.1213} {\bibfield
  {journal} {\bibinfo  {journal} {Rev. Mod. Phys.}\ }\textbf {\bibinfo {volume}
  {87}},\ \bibinfo {pages} {1213} (\bibinfo {year} {2015})}\BibitemShut
  {NoStop}%
\bibitem [{SM()}]{SM}%
  \BibitemOpen
  \href@noop {} {\bibinfo  {journal} {Supplementary Materials}\ ,\ \bibinfo
  {pages} {for details of the derivations of longitudinal resistivity from the
  diffusion equation with anisotropic spin relaxation time, which includes
  Ref.~\cite{zhang2024microscopic}.}}\BibitemShut {Stop}%
\bibitem [{\citenamefont {Zhou}\ \emph {et~al.}(2025)\citenamefont {Zhou},
  \citenamefont {Cheng}, \citenamefont {Hu}, \citenamefont {Chu}, \citenamefont
  {Bai}, \citenamefont {Han}, \citenamefont {Liu}, \citenamefont {Pan},\ and\
  \citenamefont {Song}}]{zhou2025manipulation}%
  \BibitemOpen
\bibfield  {journal} {  }\bibfield  {author} {\bibinfo {author} {\bibfnamefont
  {Z.}~\bibnamefont {Zhou}}, \bibinfo {author} {\bibfnamefont {X.}~\bibnamefont
  {Cheng}}, \bibinfo {author} {\bibfnamefont {M.}~\bibnamefont {Hu}}, \bibinfo
  {author} {\bibfnamefont {R.}~\bibnamefont {Chu}}, \bibinfo {author}
  {\bibfnamefont {H.}~\bibnamefont {Bai}}, \bibinfo {author} {\bibfnamefont
  {L.}~\bibnamefont {Han}}, \bibinfo {author} {\bibfnamefont {J.}~\bibnamefont
  {Liu}}, \bibinfo {author} {\bibfnamefont {F.}~\bibnamefont {Pan}},\ and\
  \bibinfo {author} {\bibfnamefont {C.}~\bibnamefont {Song}},\ }\bibfield
  {title} {\bibinfo {title} {Manipulation of the altermagnetic order in
  \text{CrSb} via crystal symmetry},\ }\href
  {https://doi.org/https://doi.org/10.1038/s41586-024-08436-3} {\bibfield
  {journal} {\bibinfo  {journal} {Nature}\ ,\ \bibinfo {pages} {1}} (\bibinfo
  {year} {2025})}\BibitemShut {NoStop}%
\bibitem [{\citenamefont {Baltz}\ \emph {et~al.}(2018)\citenamefont {Baltz},
  \citenamefont {Manchon}, \citenamefont {Tsoi}, \citenamefont {Moriyama},
  \citenamefont {Ono},\ and\ \citenamefont
  {Tserkovnyak}}]{baltz2018antiferromagnetic}%
  \BibitemOpen
  \bibfield  {author} {\bibinfo {author} {\bibfnamefont {V.}~\bibnamefont
  {Baltz}}, \bibinfo {author} {\bibfnamefont {A.}~\bibnamefont {Manchon}},
  \bibinfo {author} {\bibfnamefont {M.}~\bibnamefont {Tsoi}}, \bibinfo {author}
  {\bibfnamefont {T.}~\bibnamefont {Moriyama}}, \bibinfo {author}
  {\bibfnamefont {T.}~\bibnamefont {Ono}},\ and\ \bibinfo {author}
  {\bibfnamefont {Y.}~\bibnamefont {Tserkovnyak}},\ }\bibfield  {title}
  {\bibinfo {title} {Antiferromagnetic spintronics},\ }\href
  {https://doi.org/10.1103/RevModPhys.90.015005} {\bibfield  {journal}
  {\bibinfo  {journal} {Rev. Mod. Phys.}\ }\textbf {\bibinfo {volume} {90}},\
  \bibinfo {pages} {015005} (\bibinfo {year} {2018})}\BibitemShut {NoStop}%
\bibitem [{\citenamefont {Wahl}\ \emph {et~al.}(2007)\citenamefont {Wahl},
  \citenamefont {Simon}, \citenamefont {Diekh\"oner}, \citenamefont
  {Stepanyuk}, \citenamefont {Bruno}, \citenamefont {Schneider},\ and\
  \citenamefont {Kern}}]{wahl2007exchange}%
  \BibitemOpen
  \bibfield  {author} {\bibinfo {author} {\bibfnamefont {P.}~\bibnamefont
  {Wahl}}, \bibinfo {author} {\bibfnamefont {P.}~\bibnamefont {Simon}},
  \bibinfo {author} {\bibfnamefont {L.}~\bibnamefont {Diekh\"oner}}, \bibinfo
  {author} {\bibfnamefont {V.~S.}\ \bibnamefont {Stepanyuk}}, \bibinfo {author}
  {\bibfnamefont {P.}~\bibnamefont {Bruno}}, \bibinfo {author} {\bibfnamefont
  {M.~A.}\ \bibnamefont {Schneider}},\ and\ \bibinfo {author} {\bibfnamefont
  {K.}~\bibnamefont {Kern}},\ }\bibfield  {title} {\bibinfo {title} {Exchange
  interaction between single magnetic adatoms},\ }\href
  {https://doi.org/10.1103/PhysRevLett.98.056601} {\bibfield  {journal}
  {\bibinfo  {journal} {Phys. Rev. Lett.}\ }\textbf {\bibinfo {volume} {98}},\
  \bibinfo {pages} {056601} (\bibinfo {year} {2007})}\BibitemShut {NoStop}%
\bibitem [{\citenamefont {Hiraishi}\ \emph {et~al.}(2024)\citenamefont
  {Hiraishi}, \citenamefont {Okabe}, \citenamefont {Koda}, \citenamefont
  {Kadono}, \citenamefont {Muroi}, \citenamefont {Hirai},\ and\ \citenamefont
  {Hiroi}}]{hiraishi2024nonmagnetic}%
  \BibitemOpen
  \bibfield  {author} {\bibinfo {author} {\bibfnamefont {M.}~\bibnamefont
  {Hiraishi}}, \bibinfo {author} {\bibfnamefont {H.}~\bibnamefont {Okabe}},
  \bibinfo {author} {\bibfnamefont {A.}~\bibnamefont {Koda}}, \bibinfo {author}
  {\bibfnamefont {R.}~\bibnamefont {Kadono}}, \bibinfo {author} {\bibfnamefont
  {T.}~\bibnamefont {Muroi}}, \bibinfo {author} {\bibfnamefont
  {D.}~\bibnamefont {Hirai}},\ and\ \bibinfo {author} {\bibfnamefont
  {Z.}~\bibnamefont {Hiroi}},\ }\bibfield  {title} {\bibinfo {title}
  {Nonmagnetic ground state in \text{RuO}$_{2}$ revealed by muon spin
  rotation},\ }\href {https://doi.org/10.1103/PhysRevLett.132.166702}
  {\bibfield  {journal} {\bibinfo  {journal} {Phys. Rev. Lett.}\ }\textbf
  {\bibinfo {volume} {132}},\ \bibinfo {pages} {166702} (\bibinfo {year}
  {2024})}\BibitemShut {NoStop}%
\bibitem [{\citenamefont {Liu}\ \emph {et~al.}(2024{\natexlab{b}})\citenamefont
  {Liu}, \citenamefont {Zhan}, \citenamefont {Li}, \citenamefont {Liu},
  \citenamefont {Cheng}, \citenamefont {Shi}, \citenamefont {Deng},
  \citenamefont {Zhang}, \citenamefont {Li}, \citenamefont {Ding},
  \citenamefont {Jiang}, \citenamefont {Ye}, \citenamefont {Liu}, \citenamefont
  {Jiang}, \citenamefont {Wang}, \citenamefont {Li}, \citenamefont {Xie},
  \citenamefont {Wang}, \citenamefont {Qiao}, \citenamefont {Wen},
  \citenamefont {Sun},\ and\ \citenamefont {Shen}}]{liu2024absence}%
  \BibitemOpen
  \bibfield  {author} {\bibinfo {author} {\bibfnamefont {J.}~\bibnamefont
  {Liu}}, \bibinfo {author} {\bibfnamefont {J.}~\bibnamefont {Zhan}}, \bibinfo
  {author} {\bibfnamefont {T.}~\bibnamefont {Li}}, \bibinfo {author}
  {\bibfnamefont {J.}~\bibnamefont {Liu}}, \bibinfo {author} {\bibfnamefont
  {S.}~\bibnamefont {Cheng}}, \bibinfo {author} {\bibfnamefont
  {Y.}~\bibnamefont {Shi}}, \bibinfo {author} {\bibfnamefont {L.}~\bibnamefont
  {Deng}}, \bibinfo {author} {\bibfnamefont {M.}~\bibnamefont {Zhang}},
  \bibinfo {author} {\bibfnamefont {C.}~\bibnamefont {Li}}, \bibinfo {author}
  {\bibfnamefont {J.}~\bibnamefont {Ding}}, \bibinfo {author} {\bibfnamefont
  {Q.}~\bibnamefont {Jiang}}, \bibinfo {author} {\bibfnamefont
  {M.}~\bibnamefont {Ye}}, \bibinfo {author} {\bibfnamefont {Z.}~\bibnamefont
  {Liu}}, \bibinfo {author} {\bibfnamefont {Z.}~\bibnamefont {Jiang}}, \bibinfo
  {author} {\bibfnamefont {S.}~\bibnamefont {Wang}}, \bibinfo {author}
  {\bibfnamefont {Q.}~\bibnamefont {Li}}, \bibinfo {author} {\bibfnamefont
  {Y.}~\bibnamefont {Xie}}, \bibinfo {author} {\bibfnamefont {Y.}~\bibnamefont
  {Wang}}, \bibinfo {author} {\bibfnamefont {S.}~\bibnamefont {Qiao}}, \bibinfo
  {author} {\bibfnamefont {J.}~\bibnamefont {Wen}}, \bibinfo {author}
  {\bibfnamefont {Y.}~\bibnamefont {Sun}},\ and\ \bibinfo {author}
  {\bibfnamefont {D.}~\bibnamefont {Shen}},\ }\bibfield  {title} {\bibinfo
  {title} {Absence of altermagnetic spin splitting character in rutile oxide
  \text{RuO}$_{2}$},\ }\href {https://doi.org/10.1103/PhysRevLett.133.176401}
  {\bibfield  {journal} {\bibinfo  {journal} {Phys. Rev. Lett.}\ }\textbf
  {\bibinfo {volume} {133}},\ \bibinfo {pages} {176401} (\bibinfo {year}
  {2024}{\natexlab{b}})}\BibitemShut {NoStop}%
\bibitem [{\citenamefont {Ke{\ss}ler}\ \emph {et~al.}(2024)\citenamefont
  {Ke{\ss}ler}, \citenamefont {Garcia-Gassull}, \citenamefont {Suter},
  \citenamefont {Prokscha}, \citenamefont {Salman}, \citenamefont {Khalyavin},
  \citenamefont {Manuel}, \citenamefont {Orlandi}, \citenamefont {Mazin},
  \citenamefont {Valent{\'\i}} \emph {et~al.}}]{kessler2024absence}%
  \BibitemOpen
  \bibfield  {author} {\bibinfo {author} {\bibfnamefont {P.}~\bibnamefont
  {Ke{\ss}ler}}, \bibinfo {author} {\bibfnamefont {L.}~\bibnamefont
  {Garcia-Gassull}}, \bibinfo {author} {\bibfnamefont {A.}~\bibnamefont
  {Suter}}, \bibinfo {author} {\bibfnamefont {T.}~\bibnamefont {Prokscha}},
  \bibinfo {author} {\bibfnamefont {Z.}~\bibnamefont {Salman}}, \bibinfo
  {author} {\bibfnamefont {D.}~\bibnamefont {Khalyavin}}, \bibinfo {author}
  {\bibfnamefont {P.}~\bibnamefont {Manuel}}, \bibinfo {author} {\bibfnamefont
  {F.}~\bibnamefont {Orlandi}}, \bibinfo {author} {\bibfnamefont {I.~I.}\
  \bibnamefont {Mazin}}, \bibinfo {author} {\bibfnamefont {R.}~\bibnamefont
  {Valent{\'\i}}}, \emph {et~al.},\ }\bibfield  {title} {\bibinfo {title}
  {Absence of magnetic order in \text{RuO}$_2$: insights from $\mu$ sr
  spectroscopy and neutron diffraction},\ }\href
  {https://doi.org/https://doi.org/10.1038/s44306-024-00055-y} {\bibfield
  {journal} {\bibinfo  {journal} {npj Spintronics}\ }\textbf {\bibinfo {volume}
  {2}},\ \bibinfo {pages} {50} (\bibinfo {year} {2024})}\BibitemShut {NoStop}%
\bibitem [{\citenamefont {Noh}\ \emph {et~al.}(2025)\citenamefont {Noh},
  \citenamefont {Kim}, \citenamefont {Lee}, \citenamefont {Jung}, \citenamefont
  {Seo}, \citenamefont {So}, \citenamefont {Lee}, \citenamefont {Lee},
  \citenamefont {Park}, \citenamefont {Yang} \emph
  {et~al.}}]{noh2025tunneling}%
  \BibitemOpen
  \bibfield  {author} {\bibinfo {author} {\bibfnamefont {S.}~\bibnamefont
  {Noh}}, \bibinfo {author} {\bibfnamefont {G.-H.}\ \bibnamefont {Kim}},
  \bibinfo {author} {\bibfnamefont {J.}~\bibnamefont {Lee}}, \bibinfo {author}
  {\bibfnamefont {H.}~\bibnamefont {Jung}}, \bibinfo {author} {\bibfnamefont
  {U.}~\bibnamefont {Seo}}, \bibinfo {author} {\bibfnamefont {G.}~\bibnamefont
  {So}}, \bibinfo {author} {\bibfnamefont {J.}~\bibnamefont {Lee}}, \bibinfo
  {author} {\bibfnamefont {S.}~\bibnamefont {Lee}}, \bibinfo {author}
  {\bibfnamefont {M.}~\bibnamefont {Park}}, \bibinfo {author} {\bibfnamefont
  {S.}~\bibnamefont {Yang}}, \emph {et~al.},\ }\bibfield  {title} {\bibinfo
  {title} {Tunneling magnetoresistance in altermagnetic \text{RuO}$_2$-based
  magnetic tunnel junctions},\ }\href
  {https://doi.org/10.48550/arXiv.2502.13599} {\bibfield  {journal} {\bibinfo
  {journal} {arXiv}\ } (\bibinfo {year} {2025})},\ \Eprint
  {https://arxiv.org/abs/2502.13599} {2502.13599} \BibitemShut {NoStop}%
\end{thebibliography}
\end{document}